\documentclass[10pt]{article}
\usepackage{upgreek}
\usepackage[mathscr]{euscript}
\usepackage{amsmath,environ}
\usepackage{amsfonts} 
\usepackage{amssymb}
\usepackage{mathrsfs}
\usepackage{graphicx,psfrag,color}
\usepackage{enumerate}

\usepackage{xcolor}
\definecolor{wrongultramarine}{rgb}{0.07, 0.04, 0.56}
\usepackage[colorlinks, citecolor=wrongultramarine, linkcolor=blue, backref=true]{hyperref}
\usepackage[firstinits=true, parentracker=true, 
doi=true, isbn=false, url=true, natbib=true,
eprint=false,sorting=nyt,]{biblatex}
\newcommand{\la}{\mathrm{b}}
\newcommand{\uvec}[1]{\ensuremath{\hat{\mathbf{#1}}}}
\numberwithin{equation}{section}
\DeclareMathAlphabet{\mathpzc}{OT1}{pzc}{m}{it}
\newcommand*{\bfrac}[2]{\genfrac{}{}{0pt}{}{\raisebox{-.3em}{\scriptsize$#1$}}{\raisebox{.4em}{\scriptsize$#2$}}}
\DeclareMathOperator{\su}{\psi}

\begin{document}
\title{Scattering of electronic waves in square and triangular lattice half-planes with monoatomic 
step\thanks{The support of SERB MATRICS grant MTR/2017/000013 is gratefully acknowledged.}}
\author{Basant Lal Sharma\thanks{Department of Mechanical Engineering, Indian Institute of Technology Kanpur, Kanpur, U. P. 208016, India. {E-mail:~\textsf{bls@iitk.ac.in},
 Phone: +91\,512\,259\,6173,
 Fax: +91\,512\,259\,7408}}}
 
\maketitle

\begin{abstract}
{Scattering of electronic waves in square and triangular lattice half-planes by a step on the surface is analyzed using the nearest-neighbour tight binding approximation. The changes in lattice spacing and the transfer integral between nearest-neighbor sites near the surface are ignored. A standard application of the discrete Wiener--Hopf method leads to an exact solution of the scattering problem associated with 
incidence from the `bulk'. A far-field approximation of electronic wavefunction, as well as its graphical comparison with a numerical solution, are also provided. Natural applications and possible  extensions are based on the bulk Brillouin zone for two dimensional lattice planes as well as surface energy bands for fcc crystals.}
\end{abstract}

\setcounter{section}{-1}
\section{Introduction}
The dynamics of electrons 
in crystalline materials has evoked enormous interest historically \cite{Bloch1928,Kronig1931,Korringa1947}.
Even in present times it continues to play a fundamental role at nanoscale and pose profound challenges \cite{Finzelbook,Beenakker1991,bauer2016nanoscale}; 
a phrase 
that succinctly summarizes the concomitant puzzles is `size effect' \cite{Tellier1982}. However the dimensional issues have been raised since the foundations of the electron theory of metals 
were laid \cite{Thomson1901,goodwin1939electronic1,goodwin1939electronic2,goodwin1939electronic3,goodwin1939electronic4}, 
and the high specific electrical resistance exhibited by thin metal films has been explained by the limitation of the electronic mean free path due to the geometry of the film \cite{Andrew1949,Tellier1982}. 
The surface contribution to the resistivity is attributed to surface roughness, i.e. 
the deviation from a perfectly plane surface.
Indeed, the ubiquitous surface scattering is still elusive while it is considered to play a key role in the increase in resistivity of 
thin films \cite{SamblesFilm1983,Hensel1985,Graham2010,SunYao2010,rurali2010colloquium}, see also \cite{Marom2006,Henriquez2013}.

In 1938, based on the scattering of conduction electrons at the film surfaces, supported by early experimental data, K. Fuchs \cite{fuchs1938} formulated a theory that related decrease in thickness to the increase in electrical resistivity of thin metal films \cite{Finzelbook}. 
This description of the surface is strictly phenomenological and assumes that the state of the surfaces can be adequately described by the single specularity parameter ${\mathit{p}}$ \cite{Larson1971}. Clearly, it is not detailed enough to adhere to any specific microscopic scattering mechanism. Nevertheless, Fuchs's theory is widely applied to analyse experimental data even to this date. A convenient form of this expression for the conductivity has been given by Chambers \cite{Chambers1950} and Sondheimer \cite{Sondheimer1952}.
From a theoretical point of view, several modifications of such model have been proposed in the literature \cite{Parrott1965,Mayadas1970,Namba1970,Sondheimer75,Hoffmann1985,Fishman1991,Palasantzas1997} for each component of the total resistivity; 
there have been attempts to generalize the hypothesis of a {\em single} specularity parameter ${\mathit{p}}$ \cite{Chambers1950,Sondheimer1952} to an angle-dependent specularity parameter \cite{Greene1964,Greene1966,Greene1966_3,ziman1960electrons}. A number of surface scattering mechanisms have been also analyzed
\cite{Koch1969} towards reasonable theoretical discussions of an angle-dependent specularity parameter \cite{GreeneDonell1966,Soffer1967}. 
It is well known that the geometrical surface roughness is one of the most important mechanisms on clean surfaces \cite{ziman1960electrons,Soffer1970}.
In the Fuchs-Sondheimer model \cite{fuchs1938,Sondheimer1952} 
the phenomenological parameter ${\mathit{p}} = 1$ characterizes perfectly specular electron scattering while ${\mathit{p}} = 0$ means completely diffusive scattering. In practice, the degree of specular scattering at rough surfaces is determined by fitting the ab initio data to the Fuchs-Sondheimer model, for example see \cite{Youqi2009}. 
Other more advanced analytic models have also been proposed in later years in the general area of thin-film resistivity which take into account the quantum mechanical effects that may become prominent for extremely thin films 
\cite{Trivedi1988,Fishman1991,Meyerovich2002}. 

The surface scattering of conduction electrons 
also often reduces to scattering by potential centers randomly distributed over the surface \cite{baskin1970electron,Watanabe1973,Watanabe1974}. 
In the context 
of this paper, the classic work of Greene and O'Donnell \cite{GreeneDonell1966} is relevant. 
These authors presented a detailed analysis for electron scattering from a random array of surface charges where
a crucial role is played by the differential scattering probability (calculated from the scattering amplitude for one scatterer). 
Analogous problem holds for the rough surface of crystalline materials.
\begin{figure}[ht!]
\centering{\includegraphics[width=.7\linewidth]{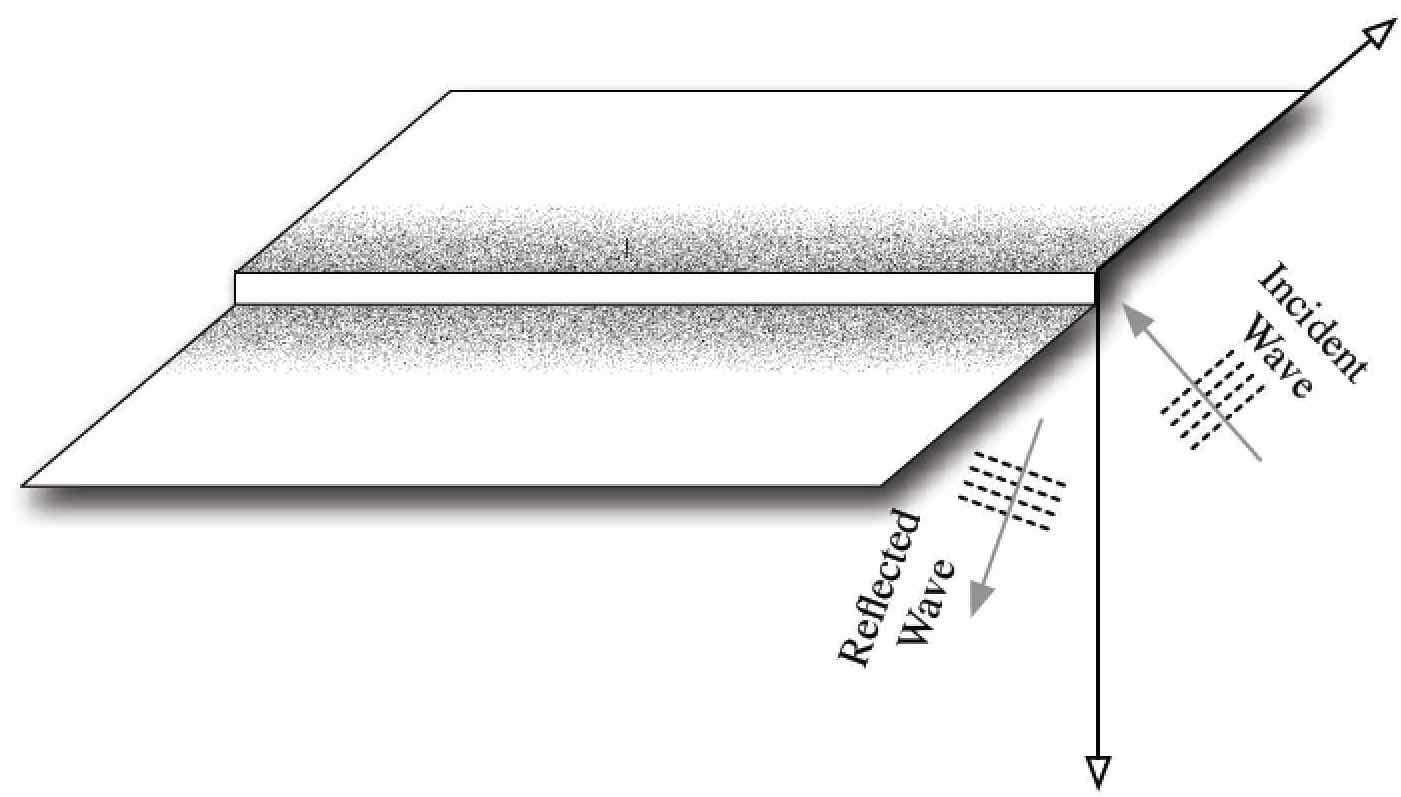}}\caption{Semi-infinite half space with step. Incident bulk (conduction) electron schematically shown. 
}\label{Fig1}
\end{figure}
Thus, in case of surface steps, as a counterpart of equation (8a) of \cite{GreeneDonell1966}, it is required that the scattered wavefunction (difference between total wavefunction and its geometric part, i.e., $\su^{{}}-\su^{{g}}$) is obtained in the far-field for a single surface step as shown in Fig. \ref{Fig1} schematically. 
This discloses the first motivation behind the present paper where it is assumed that there are no scatterers in the bulk and that the electron wavefunction vanishes at the boundary of semi-infinite lattice \cite{baskin1970electron,Andreev1972}.

A second, in a sense closely related, motivation for the paper stems from the scattering of the two-dimensional electron gas off step edges and point defects \cite{Schmid2000,matsuda2004electrical,Kurnosikov2009,biswas2011scattering,Ohmann2014}. 
The electronic structure, which is `homogeneous' on a perfect surface \cite{Koutecky1965,Schmeits1983} is, thus, perturbed by the presence of point defects or line defects such as step edges or dislocations \cite{pollmann1980defects}.
It is known that this phenomenon, for example that observed in Cu(111) \cite{crommie1993imaging}, leads to spatial oscillations (or quantum-mechanical interference patterns); the influence of bulk band states in such standing wave patterns observed in STM has been also demonstrated \cite{Pascual2006,KimYe2011}.
For instance, in Fig. 2 of \cite{Pascual2006} a few oscillations are attributed to the scattering of bulk states at monatomic step which disperse as the energy is increased.
In the same vein, the energy dependence of the contribution of bulk state electrons to the electron standing-wave patterns at the Au (111) surface have been also systematically investigated \cite{Schouteden2009}. Patterns have been observed that originate from the scattering of bulk electrons between subsurface impurities as well as those on the surface \cite{Sprodowski2010,Vazquez2009}. 
Thus, electron interference can be induced by electron scattering at surface defects \cite{Knorr2002}, step edges, adsorbates, etc, that cause scattering patterns in the form of standing waves. In addition to the surface states \cite{crommie1993imaging,Hasegawa1993,Rutter2007} or the image-potential states \cite{Wahl2003}, the bulk states are also held responsible for some of the observed standing waves \cite{Petersen1998,Pascual2006,Schouteden2009}. The presented work concerns the bulk electron scattering from edges, ignoring the role of localized edge states, and focussing on the essential role of the dependence on incoming angle and energy of electrons \cite{Dugaev2013} (see Fig. \ref{Fig3} for bands). 

\begin{figure}[ht!]\centering
{\includegraphics[width=\linewidth]{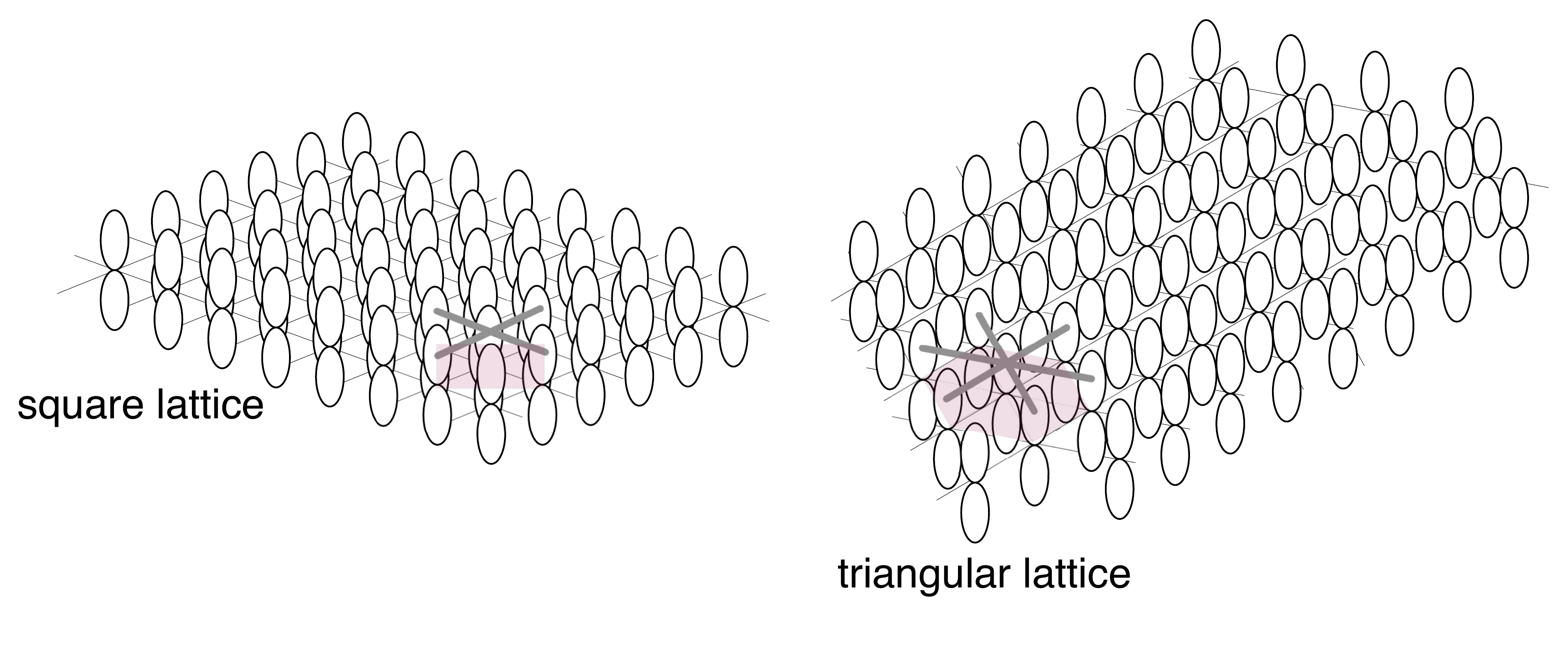}}
\caption{Schematic of square and triangular lattice structures with $p$-like orbital per atom and one atom per unit cell.}
\label{Fig2}\end{figure}

With above backdrop, in this paper the scattering of electronic waves in square and triangular lattice half-planes by a surface step has been analyzed using the nearest-neighbour {\em tight binding approximation} (with a schematic illustration in Fig. \ref{Fig2}). Electronic states of crystalline systems with surfaces, specially at low temperatures, 
have been also sought earlier using the tight-binding approximation \cite{kittel1971introduction}\footnote{In chemistry this is known as the LCAO representation (linear combination of atomic orbitals), in physics it is usually called the tight-binding representation \cite{Slater1954}. The present model is the nearest neighbor tight-binding approximation \cite{callaway1964energy} as called by physicists and among chemists it is related to what is known as the H\"{u}ckel approximation \cite{Huckel1930}. The electronic band structure is calculated using an approximate set of wavefunctions based upon superposition of orbitals located at each individual atomic site \cite{Pettiforbook}. The tight-binding approximation \cite{kittel1971introduction} is typically characterized by two terms: a kinetic term that describes the hopping of particles between neighboring sites in the potential and an on-site interaction term (assumed to be zero as it appears as an offset only). };
see its rich history as revealed by the analyses of \cite{goodwin1939electronic1,goodwin1939electronic2,goodwin1939electronic3,goodwin1939electronic4}, as well as its extensions \cite{baldock_1952,Koutecky1957,Kunne1967,Cooper1970,
davison1970,Steslicka1977}. 
An elaborate and classical discussion on several types of surfaces, interfaces, 
overlayer systems, super-lattices, defects at surfaces or interfaces, etc., is provided by \cite{Pollmann1980} within the context of the tight binding and the scattering theoretical approach for locally perturbed solids \cite{koster1954wave,Slater1954}. The present paper, however, employs an exceedingly simple formulation in comparison though the calculations are less intensive and analysis is tractable.
For the case of infinite square and triangular lattices with a finite \cite{Bls3,Bls6} or a semi-infinite \cite{Bls1,Bls4} slit, while assuming the vanishing of the electronic wavefunction at the `missing' sites, the associated scattering problems 
have been recently analyzed by \cite{Bls1,Bls3,Bls4,Bls6} using the same method \cite{Noble} as that implemented in the present paper.

\begin{figure}[ht!]\centering
{\includegraphics[width=\linewidth]{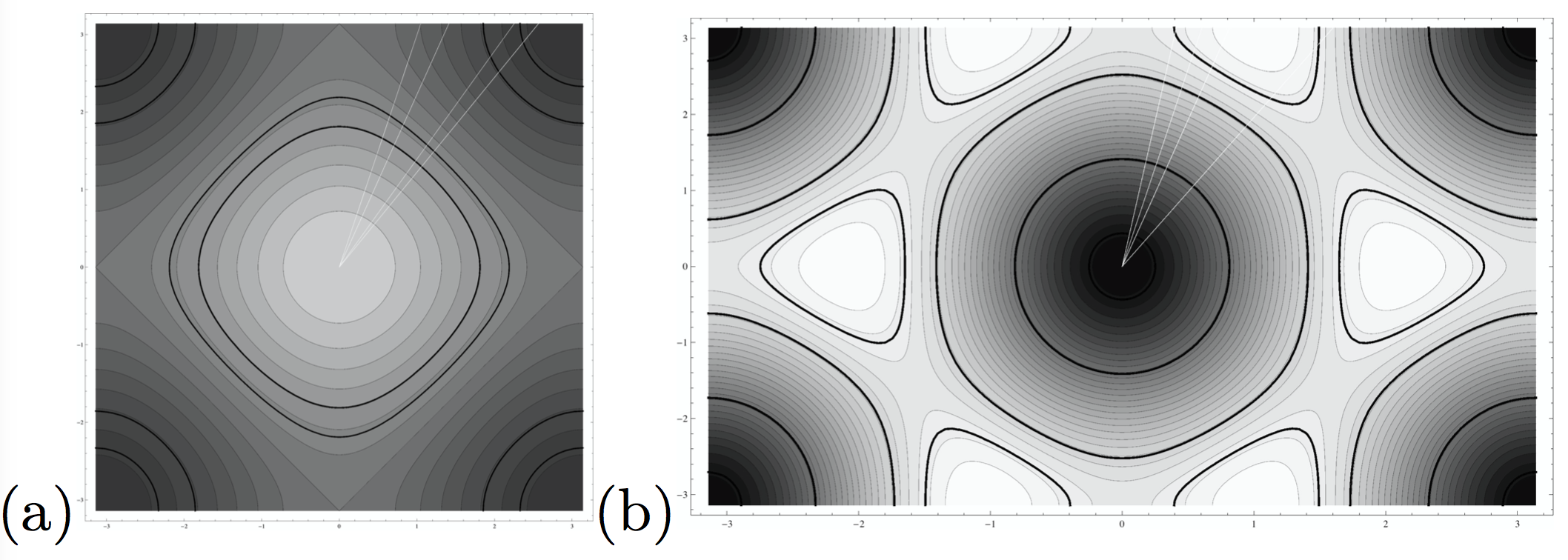}}
\caption{Energy bands ${\mathcal{E}}_{{\upkappa}}({\upkappa}_x, {\upkappa}_y)$ in the bulk for (a) ${\mathfrak{S}\hspace{-.4ex}}_{{\bullet}{\bullet}}$ and (b) ${\mathfrak{T}\hspace{-.4ex}}_{{\bullet}{\bullet}}$.}
\label{Fig3}
\end{figure}

\begin{figure}[ht!]\centering
{\includegraphics[width=\linewidth]{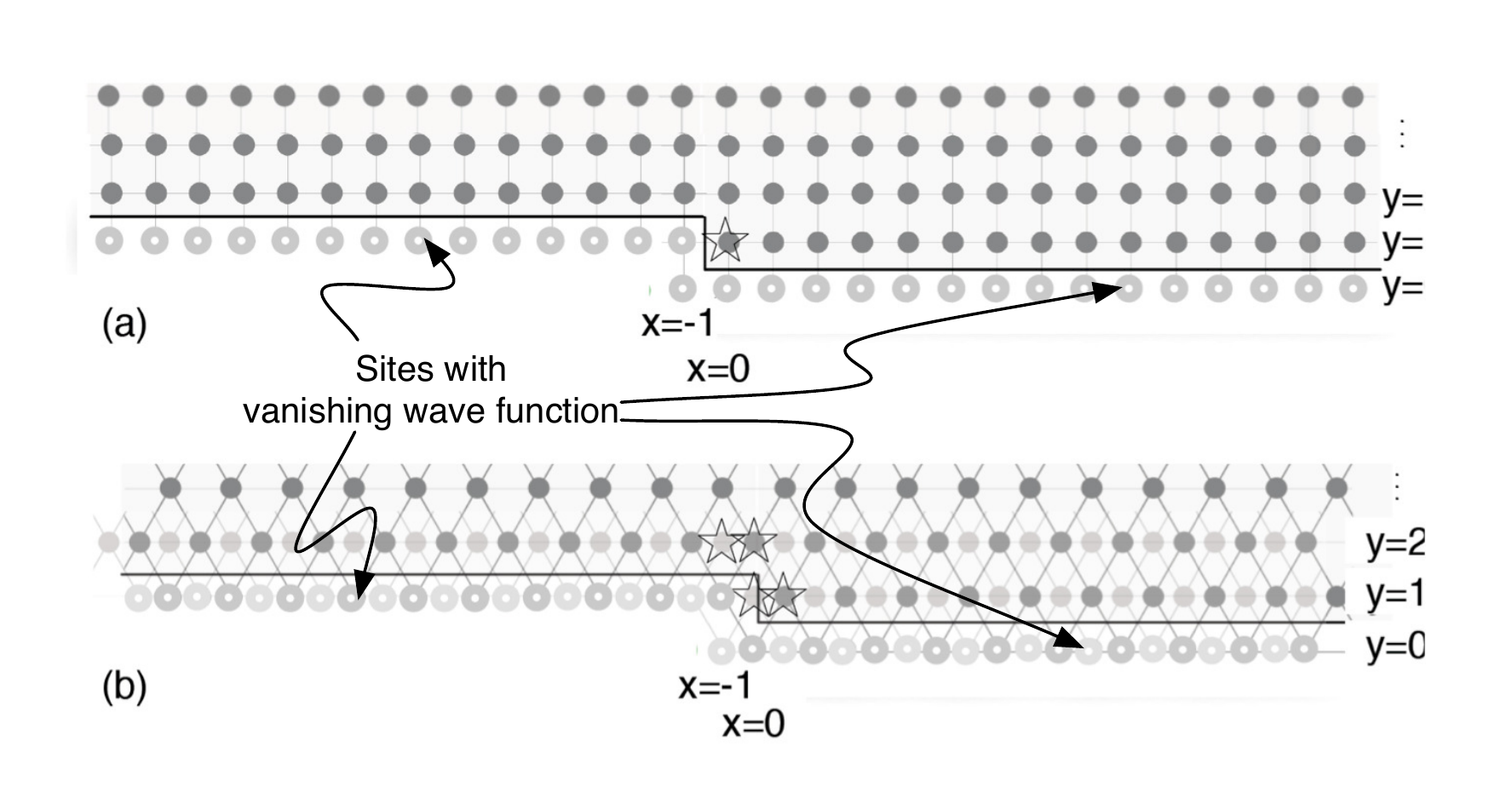}}
\caption{ Semi-infinite square and triangular lattice structure with step 
(a) ${\mathfrak{S}\hspace{-.4ex}}_{{\bullet}{\bullet}}$, (b) ${\mathfrak{T}\hspace{-.4ex}}_{{\bullet}{\bullet}}$.}
\label{Fig4}\end{figure}

\section{Square lattice model}
\label{scatter_sq}
\begin{figure}[ht!]\centering
{\includegraphics[width=\linewidth]{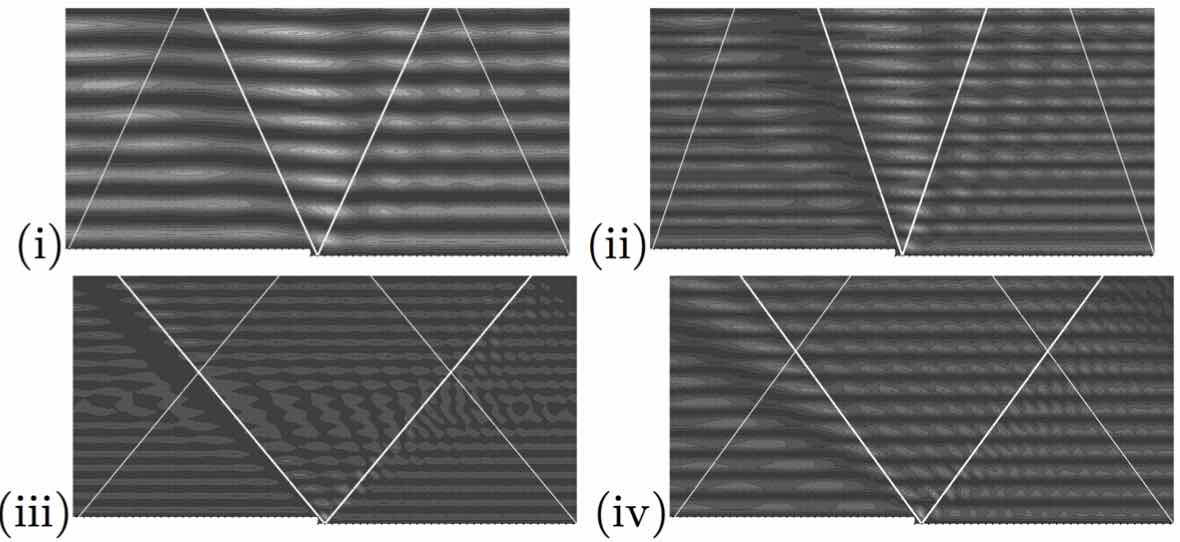}}
\caption{$|\su^{{}}|^2$ for ${\mathfrak{S}\hspace{-.4ex}}_{{\bullet}{\bullet}}$. 
(i) $\beta^{-1}{\mathcal{E}}_{{\upkappa}}=-3.38, {\upkappa}^{\mathrm{inc}}=0.81, {\Theta}=65.8$ deg, 
(ii) $\beta^{-1}{\mathcal{E}}_{{\upkappa}}=-2.56, {\upkappa}^{\mathrm{inc}}=1.27, {\Theta}=71.5$ deg, 
(iii) $\beta^{-1}{\mathcal{E}}_{{\upkappa}}=0.84, {\upkappa}^{\mathrm{inc}}=2.8, {\Theta}=50.7$ deg, and 
(iv) $\beta^{-1}{\mathcal{E}}_{{\upkappa}}=1.52, {\upkappa}^{\mathrm{inc}}=2.57, {\Theta}=54.33$ deg. 
${{\mathrm{A}}}=1, {\mathcal{E}}_2=10^{-3}, N_{\text{grid}}=71, N_{pml}=58.$}
\label{Fig5}
\end{figure}

\begin{figure}[ht!]\centering
{\includegraphics[width=.9\linewidth]{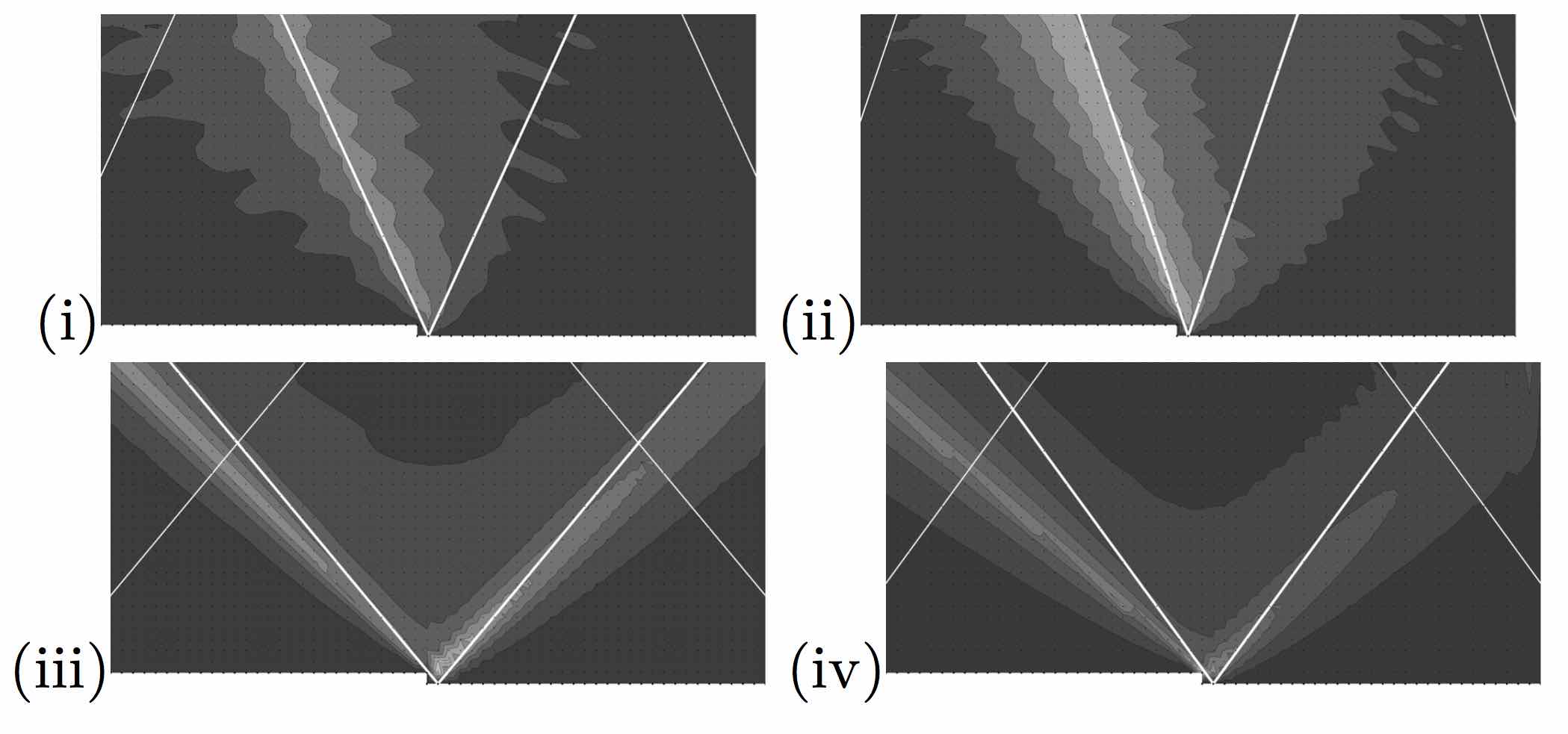}}
\caption{$|\su^{{}}-\su^{{g}}|$ for ${\mathfrak{S}\hspace{-.4ex}}_{{\bullet}{\bullet}}$. The details correspond to Fig. \ref{Fig5}.}
\label{Fig6}
\end{figure}
Let a semi-infinite two-dimensional square lattice with a surface step be denoted by ${\mathfrak{S}\hspace{-.4ex}}_{{\bullet}{\bullet}}$, i.e.,
\begin{equation}\begin{split}
{{\mathfrak{S}\hspace{-.4ex}}_{{\bullet}{\bullet}}}{:=}\{
m\la\uvec{i}+n\la\uvec{j}| 
m\in{\mathbb{Z}}, n\in{\mathbb{Z}^+}\setminus\{0,1\}\}\\
\cup\{m\la\boldsymbol{i}+\la\boldsymbol{j}| m\in{\mathbb{Z}^+}\},
\label{sqXX}
\end{split}\end{equation}
where $\uvec{i}$ and $\uvec{j}$ form the standard 
basis of two dimensional space ${\mathbb{R}}^2$.
The geometric structure of step 
in semi-infinite square lattice is schematically illustrated in Fig. \ref{Fig4}(a).

An electronic model based on the { tight-binding approximation} is considered, where 
the tunneling amplitude for an electron to hop from one atom to the next is determined by the coupling matrix element $\beta$ (i.e., the transfer integral between nearest-neighbor sites).
For the two-dimensional square lattice with either one $s$-like orbital or one $p$-like orbital per atom and one atom per unit cell, 
the tight-binding Hamiltonian 
can be explicitly written (within the second quantization) in the following form\footnote{The perturbation in lattice spacing or $\beta$ near the boundary is ignored in this simple model. {In \eqref{WSNEeqB1}, using the traditional symbolic choice, h.c. stands for the phrase `Hermitian conjugate' \cite{Surjanbook}.}}
\begin{equation}\begin{split}
{\mathcal{H}}=&-\beta\sum_{{\mathtt{x}}\in{\mathbb{Z}},{\mathtt{y}}\in{\mathbb{Z}^+}}(a_{{\mathtt{x}}+1,{\mathtt{y}}}^\dagger a_{{{\mathtt{x}}},{\mathtt{y}}}+a_{{\mathtt{x}},{\mathtt{y}}+1}^\dagger a_{{\mathtt{x}},{\mathtt{y}}}+h.c.),
\label{WSNEeqB1}
\end{split}\end{equation}
where $a^\dagger_{{\mathtt{x}},{\mathtt{y}}}$ and $a_{{\mathtt{x}},{\mathtt{y}}}$ are the {\em creation} and {\em annihilation} operators on the lattice site $({\mathtt{x}}, {\mathtt{y}})$ respectively. 
In this context, the bulk band structure of two-dimensional square lattice with one atom per unit cell \cite{kaxiras2003atomic} is easily found to be 
given by
\begin{equation}\begin{split}
{\mathcal{E}}_{{\upkappa}}={\mathcal{E}}({\upkappa}_x, {\upkappa}_y)
{:=}-\beta(2\cos{\upkappa}_x+2\cos{\upkappa}_y),\\
{\upkappa}_x\in[-\pi, \pi], {\upkappa}_y\in[0,\pi].
\label{dispersion_sq}
\end{split}\end{equation}
Applying the quantum mechanical bra-ket notation, and the Fourier transform \eqref{unpm} along ${{\mathtt{x}}}$-axis, the electronic wavefunction is expressed as
$|\Psi({{\upxi}})\rangle=\sum_{{\mathtt{y}}\in{\mathbb{Z}^+}}\su_{{{\mathtt{y}}}}({{\upxi}})\alpha_{{\upxi}}^\dagger({{\mathtt{y}}})|0\rangle,$
where $\alpha_{{\upxi}}$ (resp. $\alpha^\dagger_{{\upxi}}$) denote the Fourier transform of $a_{{\mathtt{x}},\cdot}$ (resp. $a^\dagger_{{\mathtt{x}},\cdot}$), and $|0\rangle$ denotes the vacuum wavefunction (as a reference). Then, the Schr{\"{o}}dinger equation
${\mathcal{H}}({{\upxi}})|\Psi({{\upxi}})\rangle= {\mathcal{E}}_{{\upkappa}}|\Psi({{\upxi}})\rangle,$
leads to the difference equation
\begin{equation}\begin{split}
\beta^{-1}{\mathcal{E}}_{{\upkappa}}\su_{{{\mathtt{y}}}} = -\su_{{{\mathtt{y}}}+1}({{\upxi}}) -\su_{{{\mathtt{y}}}-1}({{\upxi}})-2\cos{\upxi} \su_{{{\mathtt{y}}}}({{\upxi}}).
\label{dHelmholtz12E}
\end{split}\end{equation}
The boundary condition for ${{\mathfrak{S}\hspace{-.4ex}}_{{\bullet}{\bullet}}}$ is $\su_{{\mathtt{y}};+}({{\upxi}})|_{{\mathtt{y}}=0}=\su_{{\mathtt{y}};-}({{\upxi}})|_{{\mathtt{y}}=1}=0$ using the notation of \eqref{unpm}.

Suppose ${\su}^{\mathrm{inc}}$ describes the {\em incident electronic wave}. 
It is assumed that ${\su}^{\mathrm{inc}}$ is given by
\begin{equation}\begin{split}
{\su}_{{\mathtt{x}}, {\mathtt{y}}}^{\mathrm{inc}}{:=}{{\mathrm{A}}}e^{-i{\upkappa}_x {\mathtt{x}}-i{\upkappa}_y {\mathtt{y}}}, \quad\quad({\mathtt{x}}, {\mathtt{y}})\in{{\mathbb{Z}^2}}, 
\label{uinc_sq}
\end{split}\end{equation}
where ${{\mathrm{A}}}\in{\mathbb{C}}$ is constant. 
As stated above, the corresponding energy is $\mathcal{E}_{{\upkappa}}=\mathcal{E}({\upkappa}_x, {\upkappa}_y)$. 
The total wavefunction ${\su}^{{}}$ at an arbitrary site in ${{\mathfrak{S}\hspace{-.4ex}}_{{\bullet}{\bullet}}}$ is a sum of the incident wavefunction ${\su}^{\mathrm{inc}}$ and the scattered wavefunction ${\su}^{{\mathit{s}}}$. 
With ${\su}^{{}}_{\cdot,{\mathtt{y}}}$ as the inverse Fourier transform of $\su_{{{\mathtt{y}}}}(\cdot)$, the equation satisfied by the (discrete) wavefunction ${\su}^{{}}$ on ${{\mathfrak{S}\hspace{-.4ex}}_{{\bullet}{\bullet}}}$ is 
\begin{subequations}\begin{eqnarray}
{\su}^{{}}_{{\mathtt{x}}+1, {\mathtt{y}}}+{\su}^{{}}_{{\mathtt{x}}-1, {\mathtt{y}}}+{\su}^{{}}_{{\mathtt{x}}, {\mathtt{y}}+1}+{\su}^{{}}_{{\mathtt{x}}, {\mathtt{y}}-1}\notag\\+\beta^{-1}{\mathcal{E}}_{{\upkappa}}{\su}^{{}}_{{\mathtt{x}}, {\mathtt{y}}}=0, 
{\mathtt{x}}\in{\mathbb{Z}},{\mathtt{y}}>1\text{ or } {\mathtt{x}}\in{\mathbb{Z}^+},{\mathtt{y}}=1,
\label{dHelmholtz_sq}\\
\text{and }{\su}^{{}}_{{{\mathtt{x}}}, {0}}=0, {\mathtt{x}}\in{\mathbb{Z}^+}, {\su}^{{}}_{{{\mathtt{x}}}, {1}}=0, {\mathtt{x}}\in{\mathbb{Z}^-}.
\label{bc_sq_X}
\end{eqnarray}\label{dHelmholtz_sq_X}\end{subequations}

The wavefunction based on the numerical solution of the problem, i.e., \eqref{dHelmholtz_sq}, \eqref{bc_sq_X} with \eqref{uinc_sq}, is illustrated in Fig. \ref{Fig5}. 
The difference between the total wavefunction and the geometric wavefield (associated with specular reflection as described in Appendix \ref{appAuxgeosol}) is illustrated in Fig. \ref{Fig6}. 
Following the tradition \cite{Eco}, 
it is assumed that ${\mathcal{E}}_{{\upkappa}}\simeq{\mathcal{E}}_{{\upkappa}}+i0,$ 
with $\beta^{-1}{\mathcal{E}}_{{\upkappa}}\in[-4, 4]\setminus 
\{0, \pm4\}$ \cite{Shaban}.
Due to the absence of localized (surface) waves on the assumed structure of semi-infinite square lattice model \cite{Pollmann1980} (see also \cite{Bls9} and Appendix C of \cite{Bls9s} where the connection with one dimensional models of \cite{Louck,Wallis1957} becomes clear), there is no loss of generality in the choice of the incident wave parameters. 

{Notice that} the sites at ${\mathtt{x}}\ge0, {\mathtt{y}}=0$ as well as ${\mathtt{x}}<0, {\mathtt{y}}=1$ are assigned zero wavefunction \eqref{bc_sq_X}, as also shown in Fig. \ref{Fig4}(a) as empty dots. 
Based on the manipulations and constructions documented in \cite{Slepyanbook} and \cite{Bls0,Bls1} (see also Appendix \ref{appRecall} for a brief recollection), as well as several similar and related techniques applied by researchers few decades ago (for instance, \cite{marcus1993tight,gao2001application}), the scattered wavefunction in the lattice half-plane \eqref{sqXX} is found to be given by
\begin{equation}\begin{split}
{\su}^{{\mathit{s}}}_{{\mathtt{y}}}{}^F ={\su}^{{\mathit{s}}}{}^F_1{\lambda}^{{\mathtt{y}}-1} \label{bulk_sq}
\end{split}\end{equation}
using the definition of ${\lambda}$ stated in \eqref{lambdadef_sq} and the general solution \eqref{gensol_sq}. 
As an analogue of \eqref{dHelmholtz_sq} for ${\mathtt{y}}=1,$ $(-4-\beta^{-1}{\mathcal{E}}_{{\upkappa}}){\su}^{{}}_{{{\mathtt{x}}}, 1}=({\su}^{{}}_{{{\mathtt{x}}}-1, 1}+{\su}^{{}}_{{{\mathtt{x}}}+1, 1}+{\su}^{{}}_{{{\mathtt{x}}}, 2}+{\su}^{{}}_{{{\mathtt{x}}}, 0}-4{\su}^{{}}_{{{\mathtt{x}}}, 1}){\mathit{H}}({\mathtt{x}}),$
so that
\begin{equation}\begin{split}
&(-4-\beta^{-1}{\mathcal{E}}_{{\upkappa}}){\su}^{{\mathit{s}}}_{1;-}+(-4-\beta^{-1}{\mathcal{E}}_{{\upkappa}}){\su}^{\mathrm{inc}}_{1;-}\\
&={\su}^{{\mathit{s}}}_{-1, 1}-{{z}} {\su}^{{\mathit{s}}}_{0, 1}+({z}+{z}^{-1}+\beta^{-1}{\mathcal{E}}_{{\upkappa}}){\su}^{{\mathit{s}}}_{1;+}+{\su}^{{\mathit{s}}}_{2;+}-{\su}^{\mathrm{inc}}_{0;+}.
\label{uF_c_sq_X}
\end{split}\end{equation}
Due to their frequent appearance in the rest of the paper, 
it is convenient to introduce the definitions 
\begin{equation}\begin{split}
{{z}}_{{P}}{:=} e^{-i{\upkappa}_x}\in{\mathbb{C}}, \delta_{D+}({{z}}){:=}\sum\nolimits_{n=0}^{+\infty}{{z}}^{-n}, {{z}}\in{\mathbb{C}}\\
\delta_{D-}({{z}}){:=}\sum\nolimits_{n=-\infty}^{-1}{{z}}^{-n}, {{z}}\in{\mathbb{C}}.
\label{zPdef_sq}
\end{split}\end{equation}
In context of the well-posedness of the Wiener--Hopf problem \cite{Bls10mixed} (as described in its \S2), consider the introduction of a factor $e^{-{\mathfrak{e}}|{\mathtt{x}}|}$ in $\su^{\mathrm{inc}}$; with an implicit assumption of the limit ${\mathfrak{e}}\to0^+$.
Using \eqref{uF_c_sq_X} and \eqref{bc_sq_X}, as well as the expression ${\su}^{\mathrm{inc}}$ \eqref{uinc_sq} and ${\mathpzc{Q}}$ \eqref{q2_sq}, it is found that
\begin{subequations}\begin{eqnarray}
{{\mathpzc{w}}}-{\su}^{\mathrm{inc}}_{0;+}+{{\mathpzc{Q}}}{\su}^{{\mathit{s}}}_{1;-}&=&{{\mathpzc{Q}}}{\su}^{{\mathit{s}}}{}^F_1-{\su}^{{\mathit{s}}}_{2{;}+}, 
\label{sliteq_sq_X}\\
{\su}^{{\mathit{s}}}_{1; -}({{z}})&=&-{{\mathrm{A}}}e^{-i{\upkappa}_y}\delta_{D-}({{z}} {{z}}_{{P}}^{-1}e^{{-}{\mathfrak{e}}}), 
\label{u1n_X}\\
\text{where }{{\mathpzc{w}}}({{z}})&{:=}&{\su}^{{\mathit{s}}}_{-1, 1}-{{z}} {\su}^{{\mathit{s}}}_{0, 1}. 
\label{q2beta_X}
\end{eqnarray}\label{u0_X}\end{subequations}
After the substitution of \eqref{bulk_sq}, i.e., the expression of ${\su}^{{\mathit{s}}}{}^F_{2}$ in terms of ${\su}^{{\mathit{s}}}{}^F_{1}$
as ${\su}^{{\mathit{s}}}{}^F_{2}={\su}^{{\mathit{s}}}{}^F_{1}{\lambda}$, 
{a rearrangement of} 
\eqref{sliteq_sq_X}
and the definition of the one-sided discrete Fourier transform \eqref{unpm} leads to the Wiener--Hopf equation for ${\su}^{{\mathit{s}}}{}^F_{2}$ (i.e., ${\su}^{{\mathit{s}}}_{2;+}$ and ${\su}^{{\mathit{s}}}_{2;-}$) as
\begin{subequations}\begin{eqnarray}
{{{\mathpzc{L}}}}{\su}^{{\mathit{s}}}_{2;+}({{z}})+{\su}^{{\mathit{s}}}_{2;-}({{z}})=(1-{{{\mathpzc{L}}}}({{z}}))({{\mathpzc{w}}}({{z}})\notag\\
-{{\mathrm{A}}}{\mathpzc{Q}} e^{-i{\upkappa}_y}\delta_{D-}({{z}}{{z}}_{{P}}^{-1}e^{{-}{\mathfrak{e}}})
-{{\mathrm{A}}}\delta_{D+}({{z}}{{z}}_{{P}}^{-1}e^{{+}{\mathfrak{e}}})),\label{WHCeq_sq_X_bulkinc}\\
\text{where }
{{{\mathpzc{L}}}}
={\frac{1}{2}}(1+\frac{{\mathpzc{r}}{\mathpzc{h}}}{{\mathpzc{Q}}}),\text{ on }{\mathscr{A}}.
\label{Lgen_sq_X}
\end{eqnarray}\label{WHeqslitfull_sq_X}\end{subequations}
The symbol ${{\mathscr{A}}}$ stands for an annulus in the complex plane where the Wiener--Hopf formulation \cite{Noble} is well-posed; 
see \S2 of \cite{Bls10mixed} for the relevant mathematical analysis. 
This concludes the mathematical formulation
of the case of square lattice half-plane with step (according to the schematic
depiction of Fig. \ref{Fig4}(a)).

\begin{figure}[ht!]\centering
{\includegraphics[width=\linewidth]{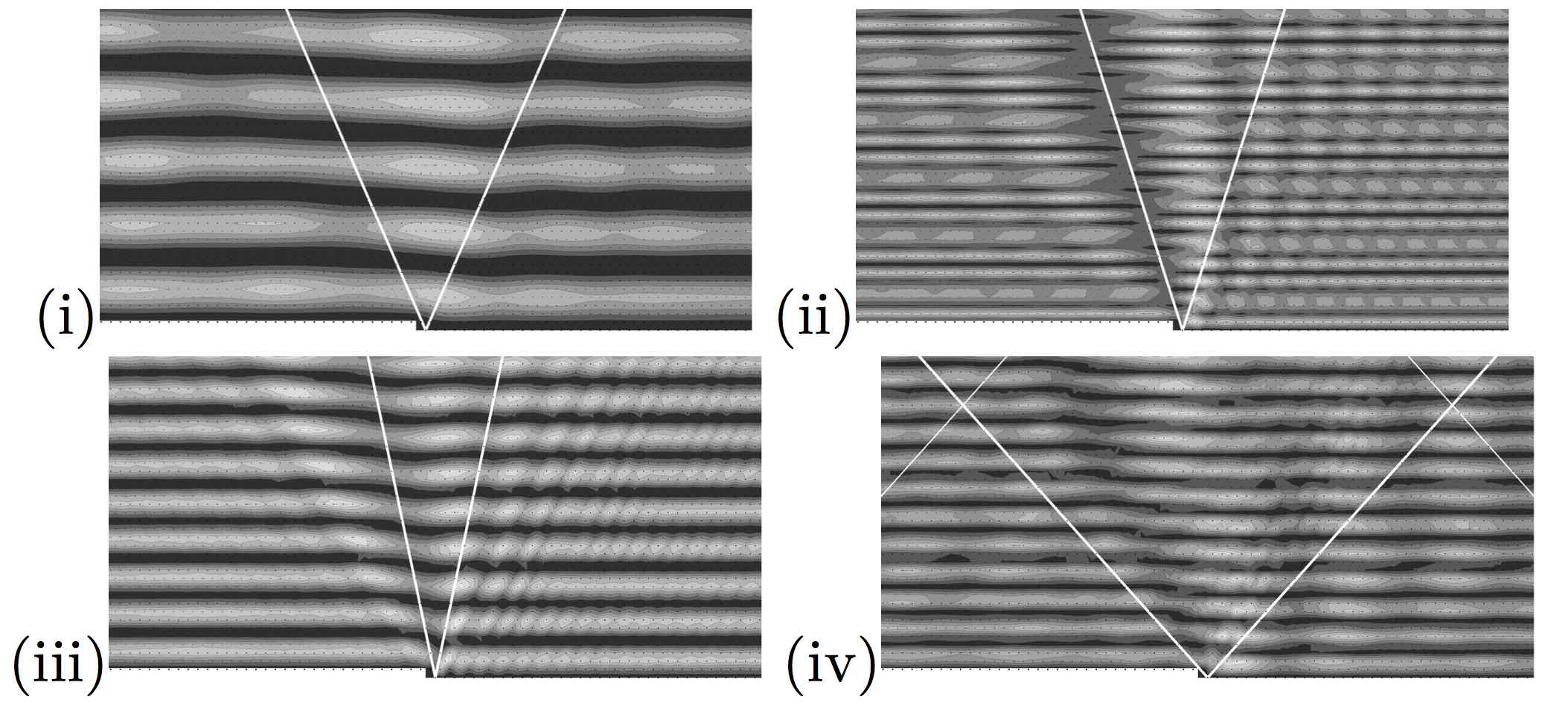}}
\caption{$|\su^{{}}|^2$ for ${\mathfrak{T}\hspace{-.4ex}}_{{\bullet}{\bullet}}$. 
(i) $\beta^{-1}{\mathcal{E}}_{{\upkappa}}=-2.75, {\upkappa}^{\mathrm{inc}}=0.5, {\Theta}=65.8$ deg, 
(ii) $\beta^{-1}{\mathcal{E}}_{{\upkappa}}=-0.75, {\upkappa}^{\mathrm{inc}}=1.63, {\Theta}=72.6$ deg, 
(iii) $\beta^{-1}{\mathcal{E}}_{{\upkappa}}=1.84, {\upkappa}^{\mathrm{inc}}=2.88, {\Theta}=78.3$ deg, and 
(iv) $\beta^{-1}{\mathcal{E}}_{{\upkappa}}=2.52, {\upkappa}^{\mathrm{inc}}=3.4, {\Theta}=48.5$ deg. 
${{\mathrm{A}}}=1, {\mathcal{E}}_2=10^{-3}, N_{\text{grid}}=101, N_{pml}=82.$ }
\label{Fig7}
\end{figure}

\begin{figure}[ht!]\centering
{\includegraphics[width=\linewidth]{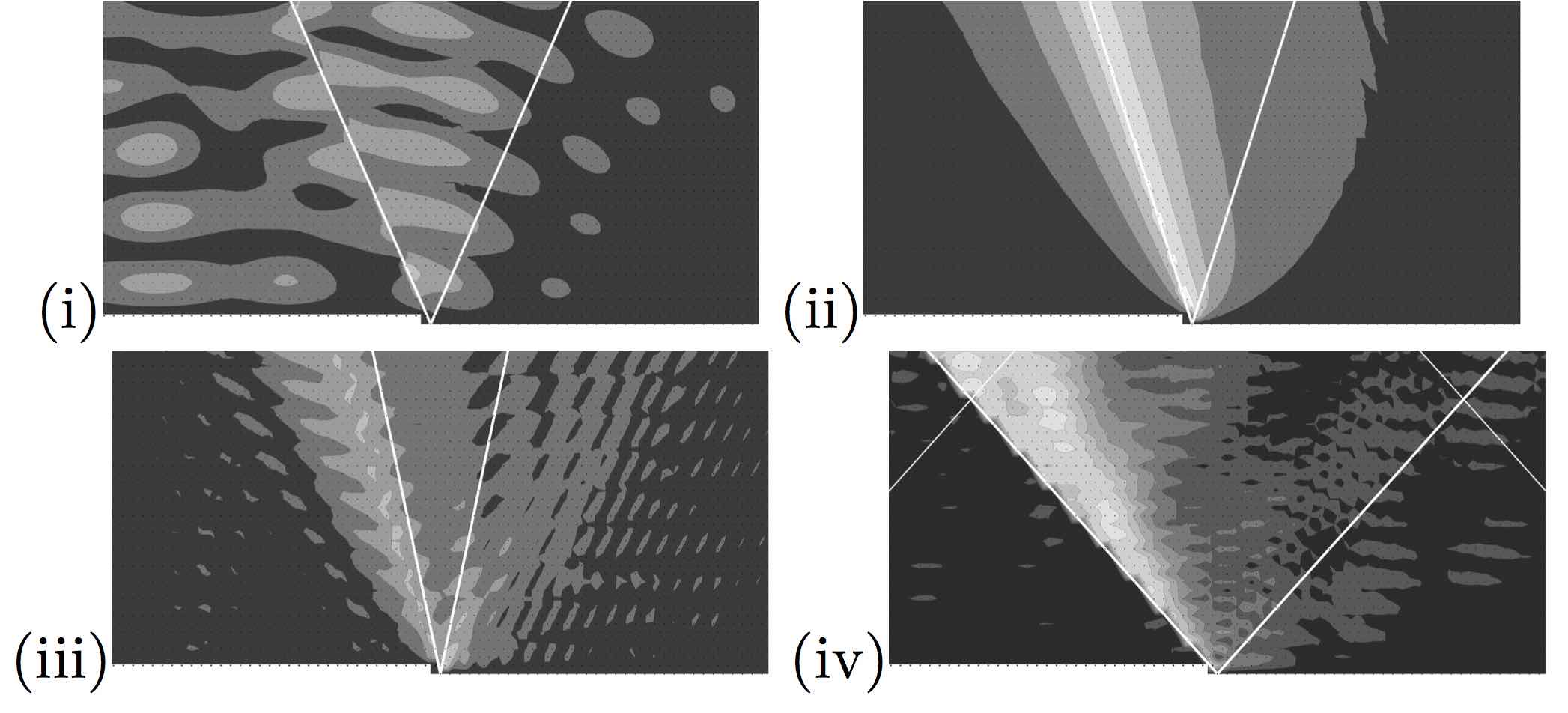}}
\caption{$|\su^{{}}-\su^{{g}}|$ for ${\mathfrak{T}\hspace{-.4ex}}_{{\bullet}{\bullet}}$. The details correspond to Fig. \ref{Fig7}.}
\label{Fig8}
\end{figure}

\section{Triangular lattice model}
\label{scatter_tg}
Let 
\begin{equation}\begin{split}
{{\mathfrak{T}\hspace{-.4ex}}_{{\bullet}{\bullet}}}{:=}\{m\boldsymbol{e}_1+n\boldsymbol{e}_2, m\in{\mathbb{Z}}, n\in{\mathbb{Z}^+}\setminus\{0,1\}\}\\
\cup\{m\boldsymbol{e}_1+\boldsymbol{e}_2, m\in{\mathbb{Z}^+}\}, \\
\text{ with }\boldsymbol{e}_1=\la\uvec{i}, \boldsymbol{e}_2={\frac{1}{2}}\la\uvec{i}+\frac{\sqrt{3}}{2}\la\uvec{j},
\label{tgXX}
\end{split}\end{equation}
represent the semi-infinite triangular lattice half-plane, as also shown schematically in Fig. \ref{Fig4}(b). 
Due to the presence of slant bonds in the triangular lattice, the formulation is placed in the sense of a union with a replicated lattice ${\mathfrak{T}\hspace{-.4ex}}_{{\bullet}{\bullet}}{{^\mathrm{R}}}$ (same as the construction introduced by \cite{Bls4}). 
This allows a rectangular coordinate system to be used in place of the usual `slant' coordinates for triangular structure. 
The union of both lattices is a rectangular lattice, denoted by ${{\mathfrak{R}}}$, with a period $\la/2$ horizontally and ${\sqrt{3}}\la/{2}$ vertically. 
The wavefunction at a site in ${{\mathfrak{R}}}$, indexed by its {\em lattice coordinates} $({\mathtt{x}}, {\mathtt{y}})\in{{\mathbb{Z}^2}}$, is denoted by ${\su}^{{\mathit{s}}}_{{\mathtt{x}}, {\mathtt{y}}}\in{\mathbb{C}}$. It is assumed that each site in ${{\mathfrak{R}}}$ is connected (atmost) with its six nearest neighbors (as part of ${\mathfrak{T}\hspace{-.4ex}}_{{\bullet}{\bullet}}$ or ${\mathfrak{T}\hspace{-.4ex}}_{{\bullet}{\bullet}}{{^\mathrm{R}}}$). 

Drawing benefit from the rectangular lattice coordinates (so that there is a symbolic similarity with the square lattice formulation presented in the previous section), it is assumed that the incident electronic wavefunction ${\su}^{\mathrm{inc}}$ is given by an expression of the form \eqref{uinc_sq},
where ${{\mathrm{A}}}\in{\mathbb{C}}$ is constant. 
The discussion concerning the tight-binding approximation follows that provided above in \S\ref{scatter_sq}.
Using the expression of the incident wave \eqref{uinc_sq}, note that the energy band relation ${\mathcal{E}}_{{\upkappa}}=\mathcal{E}({\upkappa}_x, {\upkappa}_y)$ 
for the triangular lattice satisfies \cite{Brillouin,Horiguchi} 
\begin{equation}\begin{split}
\frac{3}{2}(3+\beta^{-1}{\mathcal{E}}_{{\upkappa}})-6+2\cos2{{\upkappa}_x}+4\cos{{\upkappa}_x}\cos{{\upkappa}_y}=0, \\
{\upkappa}_x\in[-\pi, \pi], {\upkappa}_y\in[0,\pi].
\label{Tdispersion}
\end{split}\end{equation}
Because ${\mathfrak{T}\hspace{-.4ex}}_{{\bullet}{\bullet}}$ and ${\mathfrak{T}\hspace{-.4ex}}_{{\bullet}{\bullet}}{{^\mathrm{R}}}$ are `uncoupled', $[-\pi, \pi]^2$ is {\em not} the fundamental domain for ${\mathfrak{T}\hspace{-.4ex}}_{{\bullet}{\bullet}}$ \cite{Brillouin}; rather a hexagon shaped polygonal region, denoted by BZ$_1$ and also known as the first Brillouin zone \cite{Brillouin}, is the fundamental domain. 
Henceforth, it is assumed that $({\upkappa}_x, {\upkappa}_y)\in \text{BZ}_1\subset[-\pi, \pi]^2.$ The {lattice wave number} and the {angle of incidence} of ${\su}^{\mathrm{inc}}$ are defined by the relations
${\upkappa}_x={\frac{1}{2}}{\upkappa}\cos{\Theta}, {\upkappa}_y=\frac{\sqrt{3}}{2}{\upkappa}\sin{\Theta}, 
{\text{with }}{\upkappa}={\upkappa}_1+i{\upkappa}_2, {\upkappa}_1\ge0.$
In addition to the pass band of the bulk lattice, there does not exist a surface wave on the semi-infinite triangular lattice with a Dirichlet boundary \cite{Bls9,Pollmann1980}, hence there is no loss of generality in the choice \eqref{uinc_sq} of incident wave. 

The total wavefunction ${\su}^{{}}$, a sum of the incident wavefunction ${\su}^{\mathrm{inc}}$ and the scattered wavefunction ${\su}^{{\mathit{s}}}$, of an arbitrary site in the lattice ${{\mathfrak{R}}}$ (and, therefore, in triangular lattice ${\mathfrak{T}\hspace{-.4ex}}_{{\bullet}{\bullet}}$ or ${\mathfrak{T}\hspace{-.4ex}}_{{\bullet}{\bullet}}{{^\mathrm{R}}}$) satisfies the discrete Helmholtz equation (as a counterpart of \eqref{dHelmholtz_sq})
\begin{equation}\begin{split}
{\su}^{{}}_{{{\mathtt{x}}}+1, {{\mathtt{y}}}+1}+{\su}^{{}}_{{{\mathtt{x}}}+1, {{\mathtt{y}}}-1}+{\su}^{{}}_{{{\mathtt{x}}}-1, {{\mathtt{y}}}+1}+{\su}^{{}}_{{{\mathtt{x}}}-1, {{\mathtt{y}}}-1}\\
+{\su}^{{}}_{{{\mathtt{x}}}+2, {{\mathtt{y}}}}+{\su}^{{}}_{{{\mathtt{x}}}-2, {{\mathtt{y}}}}+\frac{3}{2}(\beta^{-1}{\mathcal{E}}_{{\upkappa}}-1){\su}^{{}}_{{\mathtt{x}}, {\mathtt{y}}}=0,\\
{\mathtt{x}}\in{\mathbb{Z}},{\mathtt{y}}>1\text{ or } {\mathtt{x}}\in{\mathbb{Z}^+},{\mathtt{y}}=1,
\label{dHelmholtz_tg}
\end{split}\end{equation}
along with the vanishing wavefunction condition \eqref{bc_sq_X}. 
The total electronic wavefunction based on the numerical solution of above problem is illustrated in Fig. \ref{Fig7} while the difference between the total wavefunction and the geometric wavefield (associated with specular reflection as described in Appendix \ref{appAuxgeosol}) is illustrated in Fig. \ref{Fig8}. 

The solution of the discrete Helmholtz equation \eqref{dHelmholtz_tg} (see Fig. \ref{Fig4}(b)), 
is \eqref{bulk_sq}, where ${\su}^{{\mathit{s}}}_1{}^F$ is unknown function modulo its `half' portion on the step discontinuity and ${{\lambda}}$ given by \eqref{lambdadef_sq} using the definition of ${\mathpzc{Q}}$ provided in \eqref{q2_tg}.
Using \eqref{bulk_sq} and the form of ${\su}^{\mathrm{inc}}$ \eqref{uinc_sq}, as well as 
\eqref{dHelmholtz_tg} for ${{\mathtt{y}}}=1, {{\mathtt{x}}}\ge0$ after application of the discrete Fourier transform, and the data ${\su}^{{}}_{{{\mathtt{x}}}, {\mathtt{y}}}=0$ for ${{\mathtt{y}}}=1, {{\mathtt{x}}}\in{\mathbb{Z}^-}$ and ${{\mathtt{y}}}=0, {{\mathtt{x}}}\in{\mathbb{Z}^+}$, it follows that (analogous to \eqref{u0_X})
\begin{subequations}
\begin{eqnarray}
({{z}}+{{z}}^{-1}){{\mathpzc{Q}}}({{z}}){\su}^{{\mathit{s}}}_{1;+}({{z}})={{\mathpzc{w}}}({{z}})+{{\mathpzc{w}}}^{\mathrm{inc}}({z})\notag\\+({{z}}+{{z}}^{-1}){\su}^{{\mathit{s}}}_{2;+}({{z}}), 
\label{sliteq_tg_X}\\
{\su}^{{\mathit{s}}}_{1; -}({{z}})=-{{\mathrm{A}}}e^{-i{\upkappa}_y}\delta_{D-}({{z}} {{z}}_{{P}}^{-1}), 
\label{u1n_tg_X}
\end{eqnarray}
where the complex functions ${{\mathpzc{w}}}$ and ${{\mathpzc{w}}}^{\mathrm{inc}}$ are 
\begin{equation}\begin{split}
{{\mathpzc{w}}}({{z}})&{:=} 
-{{z}}^2 {\su}^{{\mathit{s}}}_{0, 1}+{{z}}(-{\su}^{{\mathit{s}}}_{1, 1}-{\su}^{{\mathit{s}}}_{0, 2})\\&
+(-{{\mathrm{A}}}e^{-i{\upkappa}_y}{{z}}_{{P}}^{-2}+{\su}^{{\mathit{s}}}_{-1, 2})-{{\mathrm{A}}}e^{-i{\upkappa}_y}{{z}}_{{P}}^{-1}{{z}}^{-1},\\
{{\mathpzc{w}}}^{\mathrm{inc}}({z})&={{z}} \su^{\mathrm{inc}}_{0, 0}-\su^{\mathrm{inc}}_{-1, 0}.
\label{beta_tg_XR}
\end{split}\end{equation}
\end{subequations}
By virtue of the expression \eqref{bulk_sq}, it follows that 
${\su}^{{\mathit{s}}}_{2}{}^F={\su}^{{\mathit{s}}}_{1}{}^F{\lambda}$ holds;
further, upon substitution of the same in the equation \eqref{sliteq_tg_X}, after simplification, 
the discrete Wiener--Hopf equation for ${\su}^{{\mathit{s}}}_{2;+}$ and ${\su}^{{\mathit{s}}}_{2;-}$ is found to be (contrast with \eqref{WHCeq_sq_X_bulkinc})
\begin{equation}\begin{split}
{{{\mathpzc{L}}}}({{z}}){\su}^{{\mathit{s}}}_{2;+}({{z}})+{\su}^{{\mathit{s}}}_{2;-}({{z}})&=(1-{{{\mathpzc{L}}}}({{z}}))
(\frac{({{\mathpzc{w}}}({{z}})+{{\mathpzc{w}}}^{\mathrm{inc}}({{z}}))}{{{z}}+{{z}}^{-1}}\\
&-{\mathpzc{Q}}{{\mathrm{A}}}e^{-i{\upkappa}_y}\delta_{D-}({{z}} {{z}}_{{P}}^{-1}e^{{-}{\mathfrak{e}}})-{{\mathrm{A}}}\delta_{D+}({{z}} {{z}}_{{P}}^{-1}e^{{+}{\mathfrak{e}}})), \forall{{z}}\in{{\mathscr{A}}}, \label{WHCeq_tg_X}
\end{split}\end{equation}
where ${{{\mathpzc{L}}}}$ is given by \eqref{Lgen_sq_X}. Again, ${{\mathscr{A}}}$ is an annulus in the complex plane suitable for the Wiener--Hopf formulation \cite{Noble} (analogous to that for the case of square lattice). This completes 
the problem formulation for the triangular lattice half-plane with step as illustrated in Fig. \ref{Fig4}(b).

\section{The exact solution}
\label{crack_sq_WHsol}
After a standard application of the Wiener--Hopf technique (detailed calculations are presented in Appendix \ref{appAuxexactsol_sq} for ${\mathfrak{S}\hspace{-.4ex}}_{{\bullet}{\bullet}}$ and in Appendix \ref{appAuxexactsol_tg} for ${\mathfrak{T}\hspace{-.4ex}}_{{\bullet}{\bullet}}$) for solving both equations \eqref{WHCeq_sq_X_bulkinc} and \eqref{WHCeq_tg_X}, it is found that the function ${\su}^{{\mathit{s}}}_{1}{}^F$ can be expressed 
as
\begin{equation}\begin{split}
{\su}^{{\mathit{s}}}_{1}{}^F({{z}})=&{{\mathrm{A}}}{{z}}{{\mathpzc{K}}}({{z}})(\frac{{\mathtt{C}}_{0{B}}}{{{z}}-{{z}}_{{P}}\alpha_{{B}}}+\frac{{\mathtt{C}}_{0{A}}}{{{z}}-{{z}}_{{P}}\alpha_{{A}}}), \\
\alpha_{{A},{B}}&=e^{\mp{\mathfrak{e}}}, \\
{{\mathpzc{K}}}({{z}})&{:=}\frac{1}{(1-{z}_{{\mathpzc{q}}}^{1+\tau}{z}^{-1-\tau}){{{\mathpzc{L}}}}_{+}({{z}})},\\
{\mathtt{C}}_{0{B}}&{:=} e^{-i{\upkappa}_y}(1-{{z}}_{{\mathpzc{q}}}^{1+\tau}{{z}}_{{P}}^{-1-\tau}){{{\mathpzc{L}}}}_{+}({{z}}_{{P}}), \\
{\mathtt{C}}_{0{A}}&{:=}-({{z}}_{{P}}+{{z}}_{{P}}^{-1})^{\tau}\frac{{{{\mathpzc{L}}}}_{-}^{-1}({{z}}_{{P}})}{{{z}}^{-1-\tau}_{{\mathpzc{q}}}-{{z}}_{{P}}^{1+\tau}}, 
\label{u1zsol_gh}
\end{split}\end{equation}
for ${{z}}\in{{\mathscr{A}}}$ where the choice $\tau=0$ leads to the expression for ${\mathfrak{S}\hspace{-.4ex}}_{{\bullet}{\bullet}}$ while $\tau=1$ corresponds to ${\mathfrak{T}\hspace{-.4ex}}_{{\bullet}{\bullet}}$.

\begin{figure}[ht!]\centering
{\includegraphics[width=\linewidth]{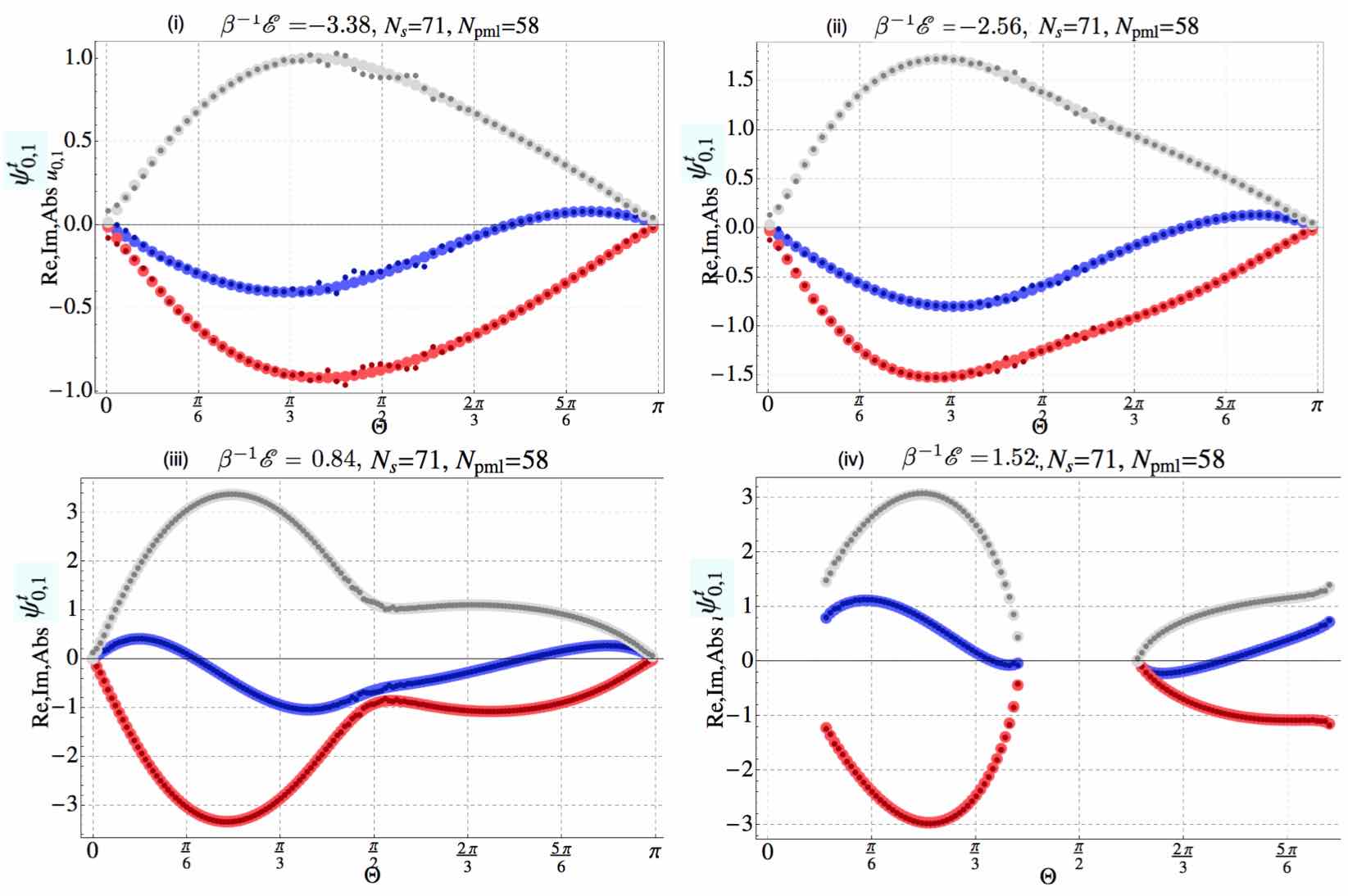}}
\caption{Re ${\su}^{{}}_{0, 1}$ (blue), Im ${\su}^{{}}_{0, 1}$ (red), and $|{\su}^{{}}_{0, 1}|$ (black) vs ${\Theta}\in(0,\pi)$ for four choices of 
$\beta^{-1}{\mathcal{E}}_{{\upkappa}}=$ (i) $-3.38,$ (ii) $-2.56$, (iii) $0.84$, (iv) $1.52$.
$N_{\text{grid}}=71, N_{\text{pml}}=58$ for ${\mathfrak{S}\hspace{-.4ex}}_{{\bullet}{\bullet}}$.
}
\label{Re_Im_Abs_u01_tot}
\end{figure}

\begin{figure}[ht!]\centering
{\includegraphics[width=\linewidth]{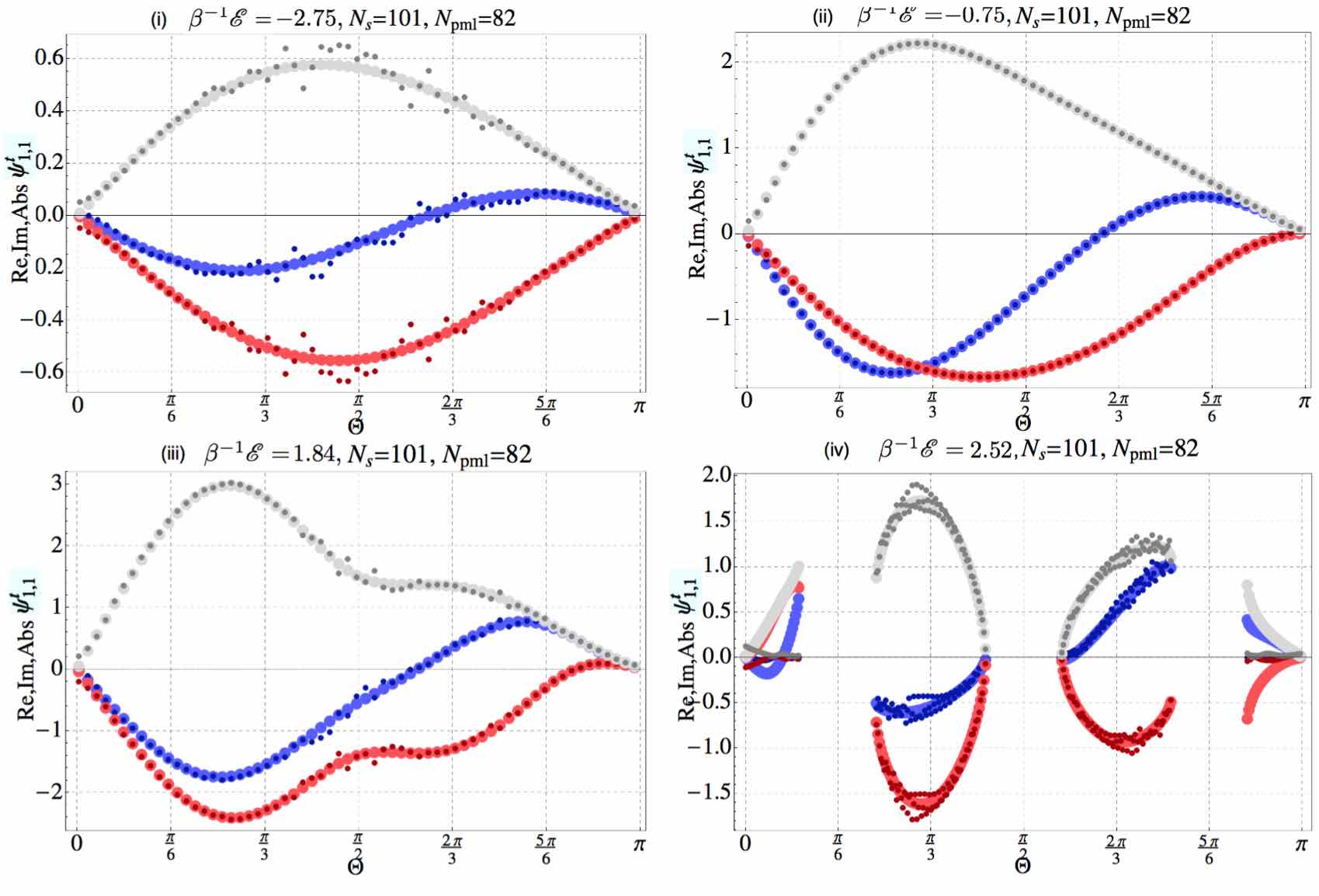}}
\caption{Re ${\su}^{{}}_{1, 1}$ (blue), Im ${\su}^{{}}_{1, 1}$ (red), and $|{\su}^{{}}_{1, 1}|$ (black) vs ${\Theta}\in(0,\pi)$ for four choices of 
$\beta^{-1}{\mathcal{E}}_{{\upkappa}}=$ (i) $-2.75,$ (ii) $-0.75$, (iii) $1.84$, (iv) $2.52$.
$N_{\text{grid}}=101, N_{\text{pml}}=82$ for ${\mathfrak{T}\hspace{-.4ex}}_{{\bullet}{\bullet}}$.
}
\label{Re_Im_Abs_u11_tot_tg}
\end{figure}

\begin{figure}[ht!]\centering
{\includegraphics[width=\linewidth]{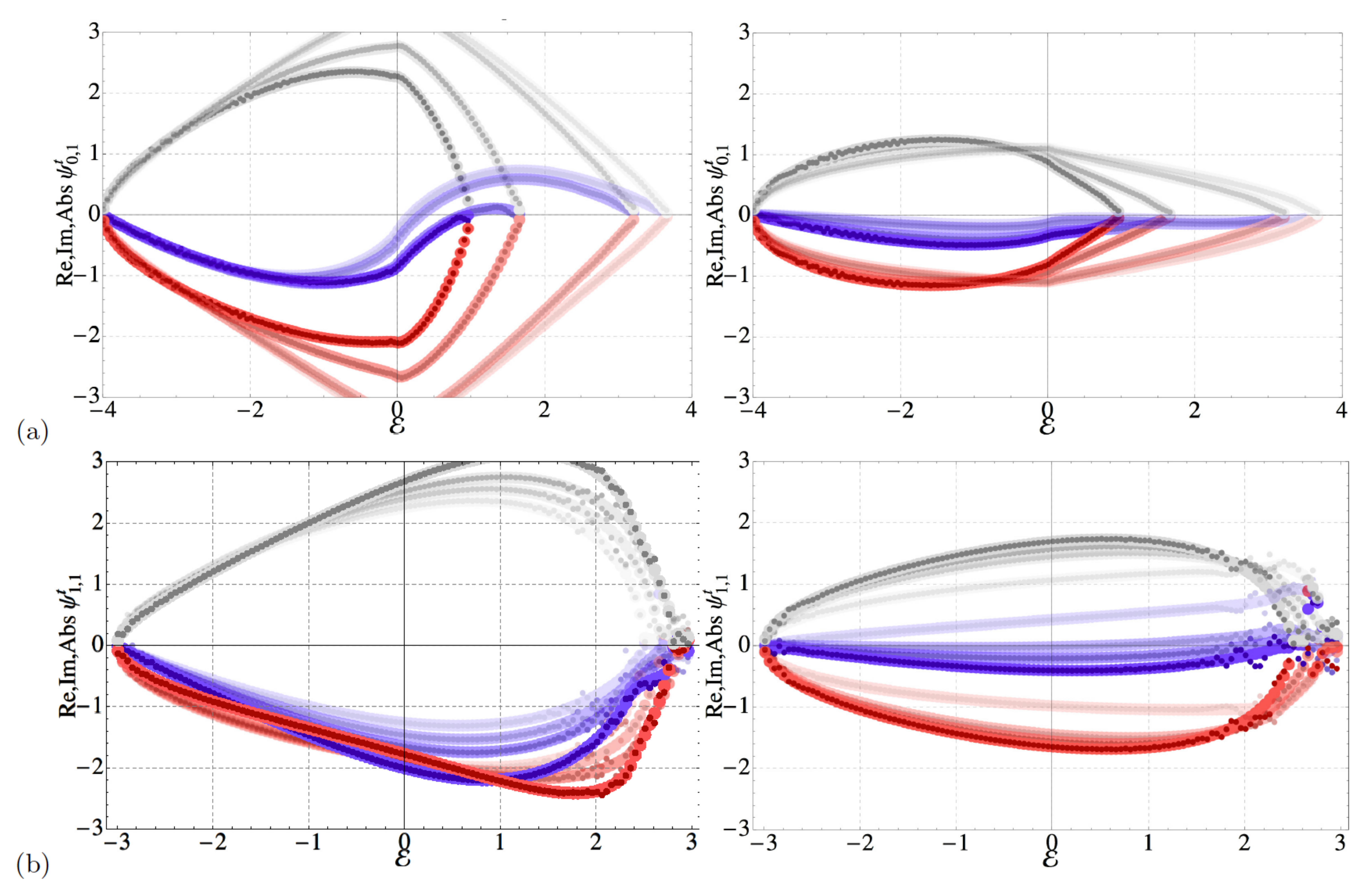}}
\caption{(a) Re ${\su}^{{}}_{0, 1}$ (blue), Im ${\su}^{{}}_{0, 1}$ (red), and $|{\su}^{{}}_{0, 1}|$ (black) vs $\beta^{-1}{\mathcal{E}}_{{\upkappa}}\in(-4,4)$ for (left) given ${\Theta}$ (right) ${\Theta}$ on ${\mathfrak{S}\hspace{-.4ex}}_{{\bullet}{\bullet}}$. The darker shades correspond to larger ${\Theta}\in\{71.52, 65.79, 54.33, 50.73\}$ (deg) on left and correspond to smaller ${\Theta}\in\{108.48, 114.21, 125.67, 129.27\}$ on right. (b) Re ${\su}^{{}}_{1, 1}$ (blue), Im ${\su}^{{}}_{1, 1}$ (red), and $|{\su}^{{}}_{1, 1}|$ (black) vs $\beta^{-1}{\mathcal{E}}_{{\upkappa}}\in(-3, 3)$ for (left) given ${\Theta}$ (right) ${\Theta}$ on ${\mathfrak{T}\hspace{-.4ex}}_{{\bullet}{\bullet}}$. The darker shades correspond to smaller ${\Theta}\in\{78.33, 72.61, 66.88, 48.54\}$ (deg) on left and 
${\Theta}\in\{101.67, 107.39, 113.12, 131.46\}$ (deg) on right. For (a) $N_{\text{grid}}=71, N_{\text{pml}}=58$, while for (b) $N_{\text{grid}}=101, N_{\text{pml}}=82$.}
\label{Re_Im_Abs_u01_tot_vsFreq_Th12_sqtg_XX}
\end{figure}

Combining \eqref{u1zsol_gh} with \eqref{bulk_sq}, ${\su}^{{\mathit{s}}}_{{\mathtt{x}}, {\mathtt{y}}}$ is eventually determined by the inverse discrete Fourier transform, 
\begin{equation}\begin{split}
{\su}^{{\mathit{s}}}_{{\mathtt{x}}, {\mathtt{y}}}=&\frac{1}{2\pi i}\oint_{{{\mathcal{C}}}}{\su}^{{\mathit{s}}}_{{\mathtt{y}}}{}^{F}({{z}}){{{z}}}^{{\mathtt{x}}-1}d{{z}}, 
\label{umnsol}
\end{split}\end{equation}
where ${{\mathcal{C}}}$ is a rectifiable, closed, counterclockwise contour in the annulus ${{\mathscr{A}}}$. 
These expressions can be simplified further in a manner akin to the results presented recently by \cite{Bls2,Bls3}. 
For example in case of ${\mathfrak{S}\hspace{-.4ex}}_{{\bullet}{\bullet}}$, using \eqref{u2pmsol_gh_sq} and \eqref{sliteq_sq_X}, the expression for ${\su}^{{\mathit{s}}}_{1;+}$ can be found 
and ${\su}^{{}}_{0, 1}$ (consequently ${\mathpzc{w}}$ by \eqref{q2beta_X}) is obtained as
\begin{equation}\begin{split}
{\su}^{{}}_{0, 1}
={{\mathrm{A}}}{l}_{+0}^{-1}({\mathtt{C}}_{0{B}}+{\mathtt{C}}_{0{A}}).
\label{u01sol_gh_sq_bulkinc}
\end{split}\end{equation}
In fact, as a rather curious observation, 
it is also found that in case of ${\mathfrak{T}\hspace{-.4ex}}_{{\bullet}{\bullet}}$ (as detailed in Appendix \ref{appAuxexactsol_tg} in order to arrive at \eqref{u01tot}), 
the same expression holds for ${\su}^{{}}_{0, 1}$ (which, incidentally, equals ${\su}^{{}}_{1, 1}$ modulo a factor $e^{-i{\upkappa}_x}$). For convenience, the corresponding sites $(0, 1)$ in ${\mathfrak{S}\hspace{-.4ex}}_{{\bullet}{\bullet}}$ and $(1, 1)$ in ${\mathfrak{T}\hspace{-.4ex}}_{{\bullet}{\bullet}}$ are marked by a star in Fig. \ref{Fig4}(a) and Fig. \ref{Fig4}(b), respectively. The graphical results are provided in Fig. \ref{Re_Im_Abs_u01_tot} for ${\su}^{{}}_{0, 1}$ in ${\mathfrak{S}\hspace{-.4ex}}_{{\bullet}{\bullet}}$ and in Fig. \ref{Re_Im_Abs_u11_tot_tg} for ${\su}^{{}}_{1, 1}$ in ${\mathfrak{T}\hspace{-.4ex}}_{{\bullet}{\bullet}}$, for various choices of incident electronic energy ${\mathcal{E}}_{{\upkappa}}$ versus the angle of incidence ${\Theta}$; Fig. \ref{Re_Im_Abs_u01_tot_vsFreq_Th12_sqtg_XX} presents the same for various choices of the angle of incidence ${\Theta}$ versus the incident electronic energy ${\mathcal{E}}_{{\upkappa}}$. The darker dots correspond to numerical solution depicted in Fig. \ref{Fig5} and Fig. \ref{Fig7}. 

Employing \eqref{umnsol}, above description provides the complete solution of the wave propagation problem in integral form. 
Notice that the `form' of the solution \eqref{umnsol} has been intentionally chosen to be the same as its counterpart for the discrete Sommerfeld problems, which were recently introduced and analyzed by the author \cite{Bls0,Bls1,Bls4}. The benefit of this choice appears below, in the form of direct application of the detailed asymptotic analysis of the scattered wavefunction in far-field 
\cite{Bls0,Bls4} (see also \cite{Bls10mixed}).

\begin{figure}[ht!]\centering
{\includegraphics[width=\linewidth]{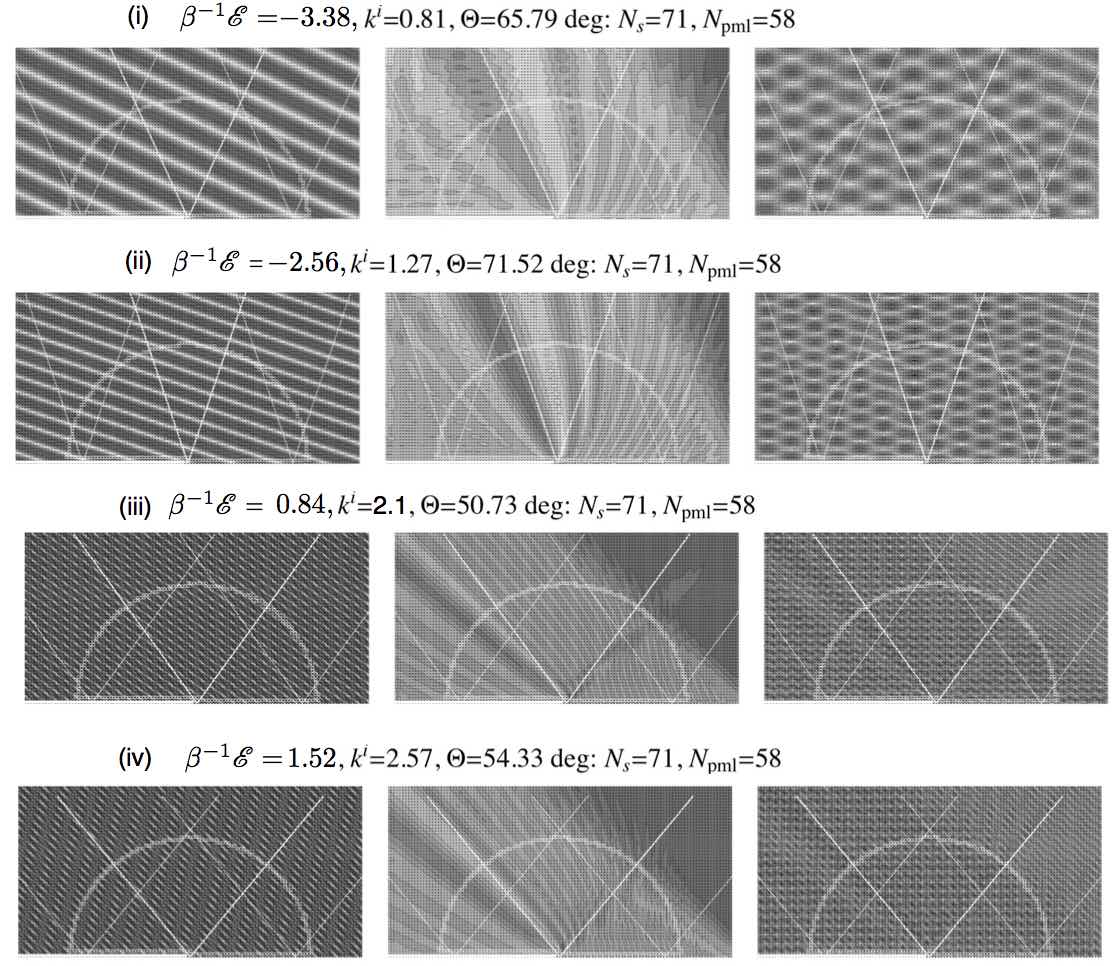}}
\caption{Incident ${\mathbb{R}}e {\su}^{\mathrm{inc}}$ (left), scattered $|{\su}^{{\mathit{s}}}|$ (center), and ${\mathbb{R}}e {\su}^{{}}$ (right) wavefunction for the semi-infinite square lattice structure with step 
${\mathfrak{S}\hspace{-.4ex}}_{{\bullet}{\bullet}}$.
{The details correspond to Fig. \ref{Fig5}.}
}
\label{stepboundary_sq_semiinfinite_num_circcurve}
\end{figure}

\begin{figure}[ht!]\centering
{\includegraphics[width=\linewidth]{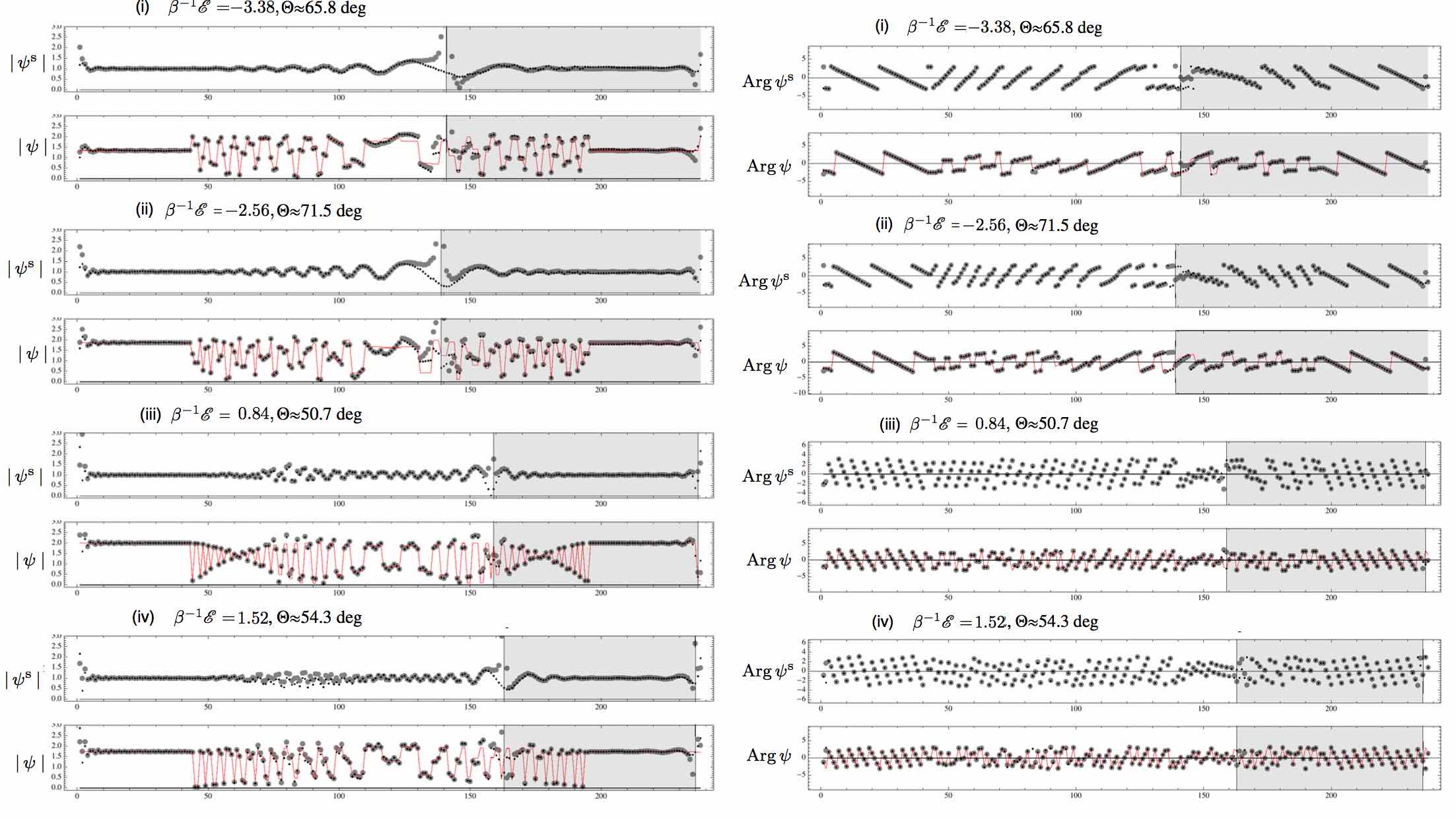}}
\caption{$|\su^{{\mathit{s}}}|$ and $|\su^{{}}|$ (left) and $\arg\su^{{\mathit{s}}}$ and $\arg\su^{{}}$ (right) on a discrete semi-circular contour as shown in Fig. \ref{stepboundary_sq_semiinfinite_num_circcurve}. In all plots, ${\mathit{R}}_\infty=40.6, N_{\text{grid}}=71, N_{\text{pml}}=58$ as stated in Fig. \ref{stepboundary_sq_semiinfinite_num_circcurve}.}
\label{AbsArgu_step_circcurve_4}
\end{figure}

\section{Far Field Approximation}
For 
\begin{equation}\begin{split}
\beta^{-1}{\mathcal{E}}_{{\upkappa}}\in\begin{cases}
(-4, 0)\cup(0, 4)&\text{ for }{\mathfrak{S}\hspace{-.4ex}}_{{\bullet}{\bullet}}\\
(-3, 7/3)\cup(7/3, 3)&\text{ for }{\mathfrak{T}\hspace{-.4ex}}_{{\bullet}{\bullet}}
\end{cases},
\end{split}\end{equation}
(recall that ${\mathcal{E}}_{{\upkappa}}\simeq{\mathcal{E}}_{{\upkappa}}+i0, {\upkappa}\simeq{\upkappa}_1+i0$) the analysis of asymptotic approximation \cite{Born,Erdelyi1,Felsen} of the scattered wavefunction in far-field 
follows after a suitable modification of the expressions stated by \cite{Bls0} and \cite{Bls4}, respectively. For example, 
${{\mathpzc{K}}}$ and ${{\mathtt{C}}}_0$ from \eqref{u1zsol_gh} are used in place of the definitions provided by \cite{Bls0,Bls4} (note that the corresponding ${\mathpzc{L}}$ is given by \eqref{Lgen_sq_X}, which does not admit explicit factors as found in case of \cite{Bls0}). 
It is found that a far-field asymptotic approximation for ${\su}^{{\mathit{s}}}$ is
\begin{subequations}
\begin{equation}\begin{split}
{\su}^{{\mathit{s}}}_{{\mathtt{x}}, {\mathtt{y}}}\sim\sum_{{S}}{\su}^{{\mathit{s}}}_{{\mathtt{x}}, {\mathtt{y}}}|_{{S}}+\sum_{{s}={A},{B}}{\su}^{{\mathit{s}}}_{{\mathtt{x}}, {\mathtt{y}}}|_{{P}{s}},
\label{uapprox}
\end{split}\end{equation}
where
\begin{equation}\begin{split}
{\su}^{{\mathit{s}}}_{{\mathtt{x}}, {\mathtt{y}}}|_{{S}}\sim
-{{\mathrm{A}}}\dfrac{(1+i{\text{\rm sgn}}({\upeta}''({\upxi}_{{S}}))){{\mathpzc{K}}}({{z}}_{{S}})e^{i{{}R}\upphi({\upxi}_{{S}})}}{2\sqrt{\pi}((\frac{2}{\sqrt{3}})^{\tau}{{}R}|{\upeta}''({\upxi}_{{S}})|\sin{\theta})^{{\frac{1}{2}}}}\\(\sum_{{s}={A},{B}}\frac{{\mathtt{C}}_{0{s}}}{\alpha_{s}{{z}}_{{P}}{{z}}_{{S}}^{-1}-1})e^{-i{\upeta}({\upxi}_{{S}})},
\label{statphase}\\
\end{split}\end{equation}
and\footnote{The expression for ${\su}^{{\mathit{s}}}_{{\mathtt{x}}, {\mathtt{y}}}|_{{P}{s}}$, the residue contribution of the pole at ${{z}}_{{P}}$ is obtained after several manipulations which are omitted.}
\begin{equation}\begin{split}
{\su}^{{\mathit{s}}}_{{\mathtt{x}}, {\mathtt{y}}}|_{{P}{B}}={\su}_{{\mathtt{x}}, {\mathtt{y}}}^{{r}{B}}{{{\mathit{H}}}({\theta}-{\theta}_{{r}})}, {\su}^{{\mathit{s}}}_{{\mathtt{x}}, {\mathtt{y}}}|_{{P}{A}}=-{\su}_{{\mathtt{x}}, {\mathtt{y}}}^{{r}{A}}{{{\mathit{H}}}({\theta}_{{r}}-{\theta})},
\label{umnpole}
\end{split}\end{equation}
\end{subequations}
with ${\upeta}$ and $\upphi$ defined in \eqref{phi_sq} and \eqref{phi_tg}, ${{}R}$ and ${\theta}$ defined in \eqref{polar_sq} and \eqref{polar_tg}, and ${\mathpzc{K}}, {\mathtt{C}}_{0{{A}}}, {\mathtt{C}}_{0{{B}}}$ given in \eqref{u1zsol_gh} while the saddle point ${{z}}_{{S}}$ (${{z}}_{{S}}=e^{-i{\upxi}_{{S}}}$) is described in \cite{Bls0,Bls4} (also discussed briefly in Appendix \ref{appAuxdetail}).
Recall that $\tau=0$ for ${\mathfrak{S}\hspace{-.4ex}}_{{\bullet}{\bullet}}$ while $\tau=1$ for ${\mathfrak{T}\hspace{-.4ex}}_{{\bullet}{\bullet}}$.
For the case of ${\mathfrak{S}\hspace{-.4ex}}_{{\bullet}{\bullet}}$, there is only one saddle point and
${\theta}_{{r}}$ is defined such that for 
${\upxi}_{{S}}=
{\upkappa}_x
\text{ when }{\theta}={\theta}_{{r}}.$
For the case of ${\mathfrak{T}\hspace{-.4ex}}_{{\bullet}{\bullet}}$,
\begin{figure}[ht!]\centering
{\includegraphics[width=\linewidth]{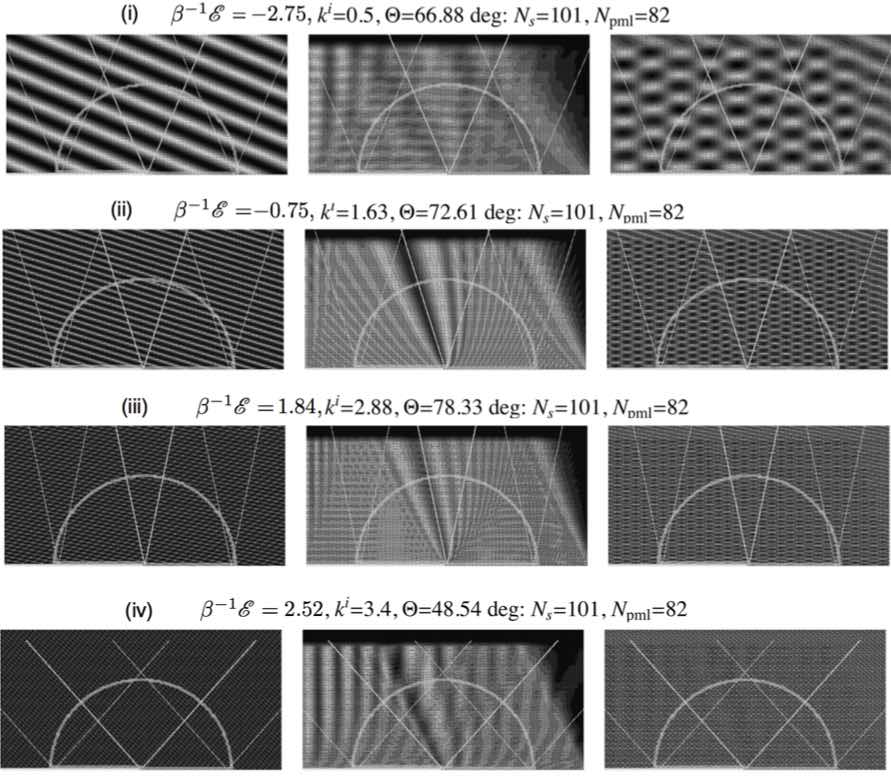}}
\caption{Incident ${\mathbb{R}}e {\su}^{\mathrm{inc}}$ (left), scattered $|{\su}^{{\mathit{s}}}|$ (center), and ${\mathbb{R}}e {\su}^{{}}$ (right) wavefunction for the semi-infinite triangular lattice structure with step 
${\mathfrak{T}\hspace{-.4ex}}_{{\bullet}{\bullet}}$.}
\label{stepboundary_tg_semiinfinite_num_circcurve_tg}
\end{figure}
\begin{figure}[ht!]\centering
{\includegraphics[width=\linewidth]{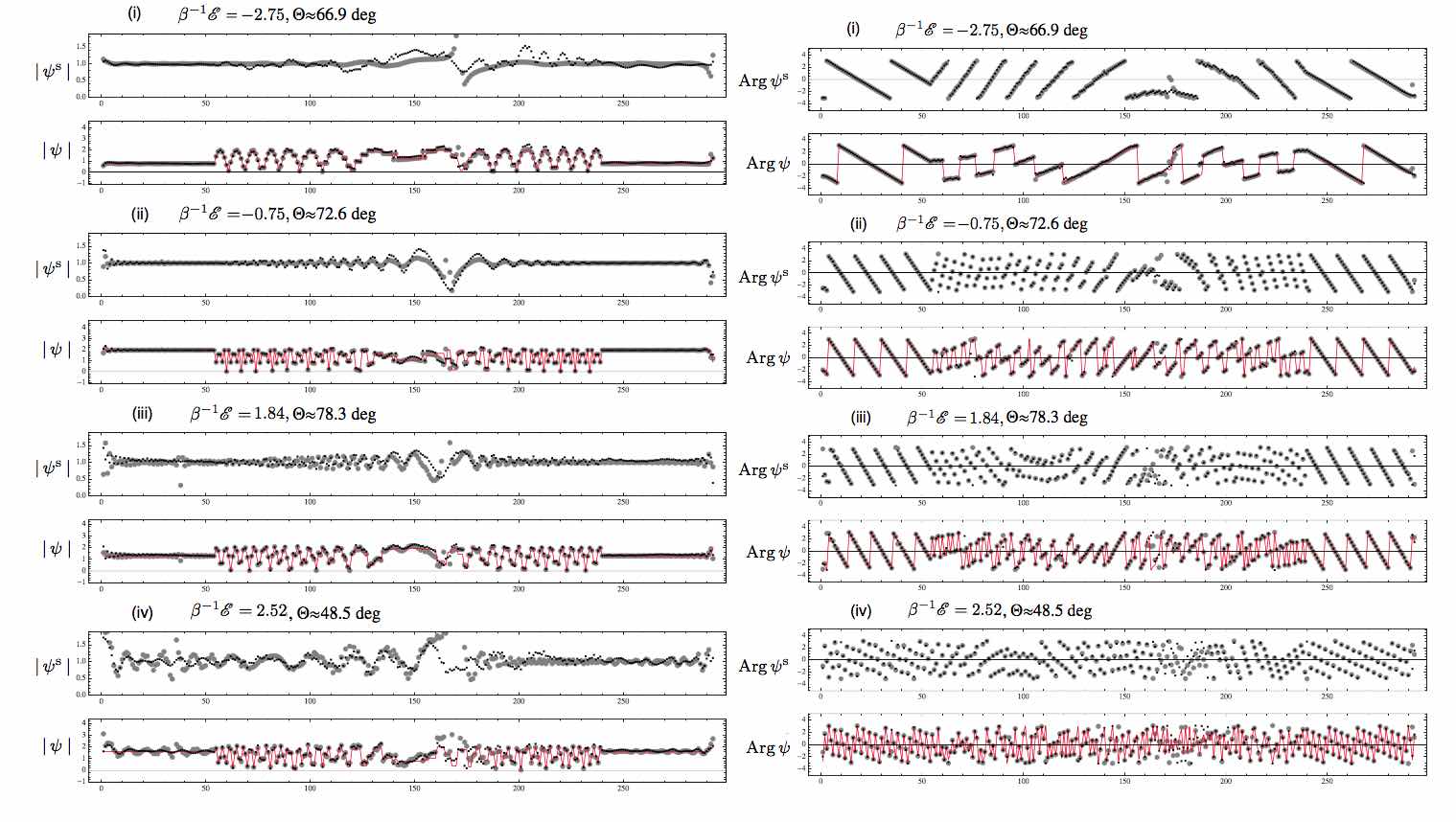}}
\caption{$|\su^{{\mathit{s}}}|$ and $|\su^{{}}|$ (left) and $\arg\su^{{\mathit{s}}}$ and $\arg\su^{{}}$ (right) on a discrete semi-circular contour as shown in Fig. \ref{stepboundary_tg_semiinfinite_num_circcurve_tg}. In all plots, ${\mathit{R}}_\infty=53.3, N_{\text{grid}}=101, N_{\text{pml}}=82$ as shown in Fig. \ref{stepboundary_tg_semiinfinite_num_circcurve_tg}.}
\label{AbsArgu_step_circcurve_tg_XX}
\end{figure}
{the notation $\sum_{{S}}$ denotes the sum over all saddle points 
of the diffraction integral that are located inside the fundamental domain of ${{\mathfrak{R}}}$ above},
and ${\theta}_{{r}}$ is defined such that for ${\theta}={\theta}_{{r}}$, ${\upxi}_{{S}}=
{\upkappa}_x$ for $\beta^{-1}{\mathcal{E}}_{{\upkappa}}\in(-3, {7}/{{3}}),$ while ${\upxi}_{{S};l}=
{\upkappa}_x$ {\em or } ${\upxi}_{{S};r}=
{\upkappa}_x$ for $\beta^{-1}{\mathcal{E}}_{{\upkappa}}\in({7}/{{3}}, {3})$. For more details concerning the saddle point analysis for the relevant diffraction integral for triangular lattice structure, see \cite{Bls4}. 

\section{Numerical Results}
Since the equations \eqref{dHelmholtz_tg}, and other equations corresponding to the defect, are algebraic, the numerical solution on a $(2N_{\text{grid}}+1)\times N_{\text{grid}}$ square grid $\Omega$ (mapped to its appropriate counterpart in case of triangular lattice structure) is straightforward. A variant of perfectly matched layers (PML) \cite{Berenger} is adopted for simulation of an `infinite' domain (see also \cite{Bls0} and \cite{Bls4}). The probability density $|\su^{{}}|^2$ is plotted in Fig. \ref{Fig5} and Fig. \ref{Fig7} for ${\mathfrak{S}\hspace{-.4ex}}_{{\bullet}{\bullet}}$ and ${\mathfrak{T}\hspace{-.4ex}}_{{\bullet}{\bullet}}$, respectively. 
Similar results are also provided in Fig. \ref{stepboundary_sq_semiinfinite_num_circcurve} and Fig. \ref{stepboundary_tg_semiinfinite_num_circcurve_tg} for ${\mathfrak{S}\hspace{-.4ex}}_{{\bullet}{\bullet}}$ and ${\mathfrak{T}\hspace{-.4ex}}_{{\bullet}{\bullet}}$, respectively.
The numerical solution, displayed in Fig. \ref{stepboundary_sq_semiinfinite_num_circcurve} and Fig. \ref{stepboundary_tg_semiinfinite_num_circcurve_tg}, is also compared with the asymptotic approximation \eqref{uapprox} of \eqref{umnsol} for ${\mathfrak{S}\hspace{-.4ex}}_{{\bullet}{\bullet}}$ and ${\mathfrak{T}\hspace{-.4ex}}_{{\bullet}{\bullet}}$, respectively. The modulus and argument of wavefunction at every site located on a (fixed) circular contour (as shown in Fig. \ref{stepboundary_sq_semiinfinite_num_circcurve} and Fig. \ref{stepboundary_tg_semiinfinite_num_circcurve_tg}) has been calculated (traversed counter-clockwise from $(0, 1)$ labelled $1$ and other sites labeled incrementally). These respective results are shown in Fig. \ref{AbsArgu_step_circcurve_4} and Fig. \ref{AbsArgu_step_circcurve_tg_XX}.
As expected based on the {low energy approximation},
since the kernel \eqref{Lgen_sq_X} approaches $1$, a `flat' surface (implying only the geometric aspect of specular scattering) behaviour is confirmed by parts (i) of Fig. \ref{AbsArgu_step_circcurve_4} and Fig. \ref{AbsArgu_step_circcurve_tg_XX} (the slight deviations are attributed to imperfect absorbing boundary on the finite grid).

\section{Discussion}
In the spirit of earlier works \cite{ziman1960electrons,Soffer1967,GreeneDonell1966,MoreLessie1973,Watanabe1973,Watanabe1974,Watanabe1976,Lessie1979,Sinha1988,Lenk1993} on electron scattering from rough surface in metals, the analysis presented in this paper can be applied in practical situations by taking appropriate convolution using the differential scattering cross section to obtain a general specularity parameter. 
A simple model such as that analyzed in this paper, leads to an exact expression for the dependence of the scattered wavefunction on the incident wavenumber and angle of incidence of the bulk electron. Such analysis is anticipated to be crucial in a theoretical framework for rough surfaces containing either a random distribution of steps or a specific structure of steps \cite{Lu1982}. For instance, the results of this type are hidden behind the extensive analyses based on careful application of statistical/variational methods since several decades for scattering from a statistically rough surface \cite{MaradudinPRB,Bass1979,Spadacini1983,Bird1985,ogilvy1991theory} as well as analytical/numerical approximations tackling multiple scattering problems \cite{garcia1979exact,shen1980multiple,desanto1986analytical}. For a finite number of steps, the geometrical ray approximation of \cite{Keller1962,Keller1957_I,Keller1957_II,keller2016rays} is reckoned highly pertinent.

An essential assumption in the paper is 
that the electron wavefunction vanishes at the metal boundary \cite{Andreev1972}, which corresponds to the approximation of the surface potential by a rectangular barrier of infinite height. 
The actual finiteness of the height and region of variation of the surface barrier affects the probability of electron scattering, leading to a smooth decay of the wavefunction within some layer near the surface. To account for the related effects, and to generalize the results of the present paper to more complex types of surface scattering is an open problem at this stage.

\begin{figure}[ht!]\centering
{\includegraphics[width=.8\linewidth]{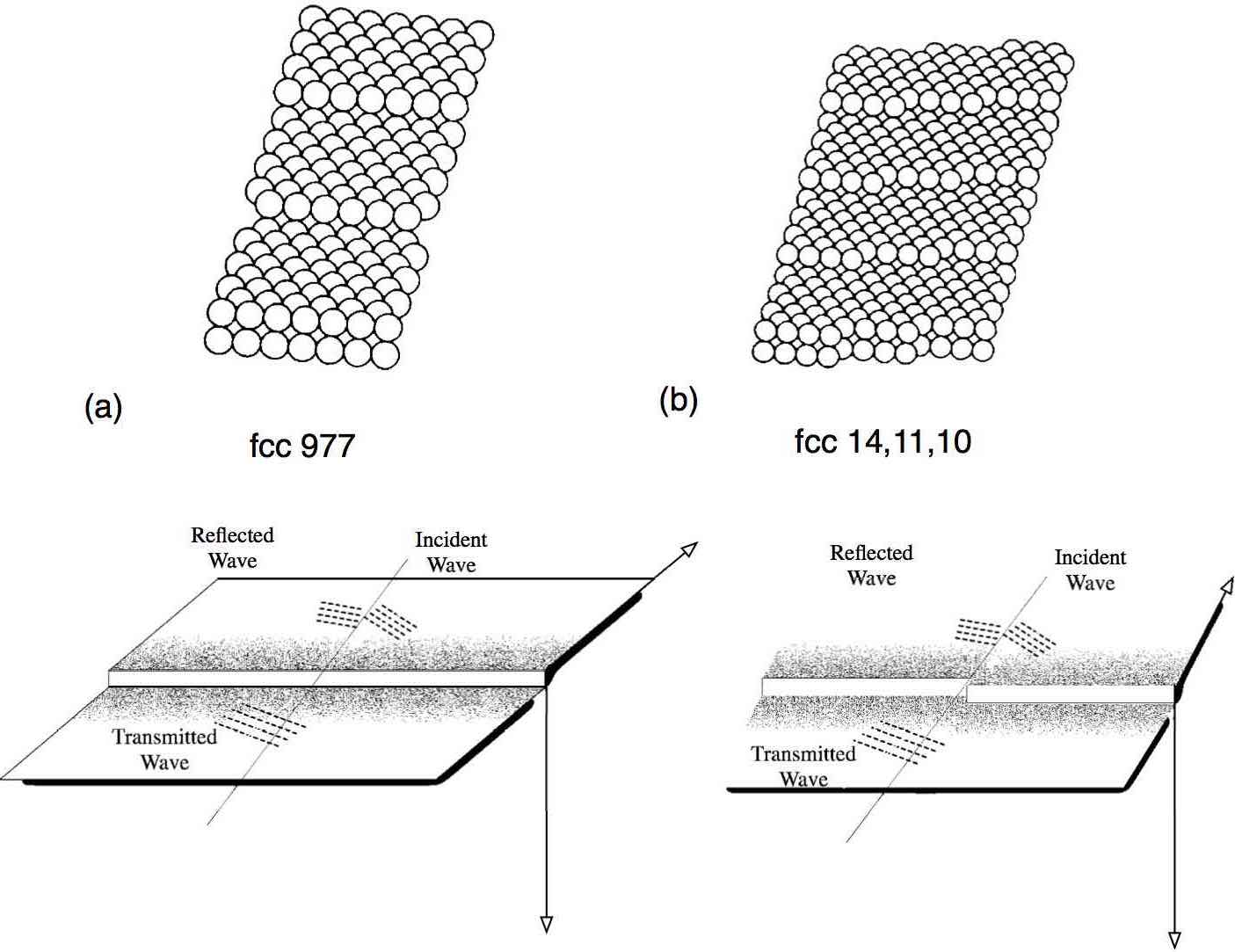}}
\caption{Semi-infinite half space with step {\cite{somorjai2010introduction}.} Incident 
surface state electron schematically shown. 
}
\label{steponhalfspace_surface}\end{figure}

The evolution of wave packets across steps on surfaces also involves inherently several challenges in a more general framework \cite{Miroshnichenko2010,cuevas2011spread}.
For surface electronic states as well, the analysis has an additional relevance based on the occurrence of surface bands, but more realistic model is wanting, hence this is deferred for another study in future. If the crystal surface is misoriented from the low-index plane by a small angle, on the atomic scale, such a surface, called a stepped or vicinal surface, is composed conventionally of terraces separated by steps of monatomic height, which may also have kinks in them (Fig. \ref{steponhalfspace_surface} top and schematic below for a formulation similar to this paper). 
\cite{Sanchez1995} demonstrated that monoatomic steps on vicinal Cu(111) surfaces act as repulsive barriers for free-electron-like, 2D surface states. The generalization of the results presented in this paper to such situation is an interesting vista that needs to be explored. 

Last but not the last the exact solution presented serves some recent interests hinged on the possible analytic solutions for quantum wells using the discrete Schr\"{o}dinger equation \cite{Boykin2004}, tight-binding models on semi-infinite lattices \cite{Tapilin1988,Skriver1991,ou1993surface,Tapilin1995}, Bloch wave scattering \cite{Morgan1966,Ziman1971,Newton1991,Wang1995}, as well as discrete nature of several interesting wave phenomena at nanoscale \cite{eisenberg1998discrete,pertsch2002anomalous,ozbay2006plasmonics}.

\section{Concluding remarks}
\label{concl}

Surface scattering is conventionally described by the well documented theoretical framework 
developed by Fuchs \cite{fuchs1938} and Sondheimer \cite{Sondheimer1952}, and later modified (for example, 
\cite{Rossnagel2004}) in different contexts to take into account the effects of surface roughness. 
This paper attempts to understand the same from a simplistic viewpoint, based on a tight-binding approximation, and presents an analysis of 
electronic wave scattering by an atomic step discontinuity \cite{Busse2000}
on the boundary of the square lattice half-plane and triangular lattice half-plane. The mathematical problem and technique is similar to that employed by \cite{Bls10mixed} which are presented in the context of physically different problem.
The extension to wider steps (multiple atomic layers) 
brings many more complicated aspects of the mathematical analysis, besides an anticipated reliance on numerics to a much greater extent. In the context of modern technological needs \cite{adams2003charge}, 
understanding the role of defects in the transport properties of graphene is central to realizing future electronics based on carbon \cite{Rutter2007} and molecules \cite{tao2006electron}. Hence, an extension of the analysis presented in this paper to more realistic, and technologically relevant structures in the current scenario, such as honeycomb structure \cite{Bls5,Bls6}, has been planned and shall be presented elsewhere. The same statement holds for the analysis of corresponding two dimensional lattice waveguides \cite{Na1998} with a step discontinuity on one or both boundary following \cite{Bls9s}. 
Such detailed 
analysis of the size effect on (ballistic) electric conductivity (or resistance) based on surface (boundary) scattering for thin `ribbons' is currently under investigation and shall be presented in future elsewhere. 

\printbibliography

\appendix

\section{Auxiliary expressions}
\label{appRecall}
\subsection{}
\label{appRecall_sq}
Following \cite{Bls0}, the discrete Fourier transform \cite{jury} ${\su}^F_{{\mathtt{y}}}: {\mathbb{C}}\to{\mathbb{C}}$ of $\{{\su}_{{\mathtt{x}}, {\mathtt{y}}}\}_{{\mathtt{x}}\in{\mathbb{Z}}}$ (along the ${\mathtt{x}}$ axis) is defined by
\begin{equation}\begin{split}
{\su}^F_{{\mathtt{y}}}={\su}_{{\mathtt{y}};+}+{\su}_{{\mathtt{y}};-}, \text{ where }\\
{\su}_{{\mathtt{y}};+}({{z}})=\sum\limits_{{\mathtt{x}}=0}^{+\infty} {\su}_{{\mathtt{x}}, {\mathtt{y}}}{{z}}^{-{\mathtt{x}}}, {\su}_{{\mathtt{y}};-}({{z}})=\sum\limits_{{\mathtt{x}}=-\infty}^{-1} {\su}_{{\mathtt{x}}, {\mathtt{y}}}{{z}}^{-{\mathtt{x}}}.\label{unpm}\end{split}\end{equation}The discrete Fourier transform ${\su}^{{\mathit{s}}}_{{\mathtt{y}}}{}^F$ of the sequence $\{{\su}^{{\mathit{s}}}_{{\mathtt{x}}, {\mathtt{y}}}\}_{{\mathtt{x}}\in{\mathbb{Z}}}$ is well defined for ${\mathtt{y}}\ge1$, using which the discrete Helmholtz equation \eqref{dHelmholtz_sq}, for all ${\mathtt{y}}$ away from the boundary, 
is expressed (recollected from \cite{Bls0}) as
\begin{subequations}\begin{eqnarray}
{{\mathpzc{Q}}}{\su}^{{\mathit{s}}}_{{\mathtt{y}}}{}^F&=&{\su}^{{\mathit{s}}}_{{\mathtt{y}}+1}{}^F+{\su}^{{\mathit{s}}}_{{\mathtt{y}}-1}{}^F, \label{dHelmholtzF_sq}\\
\text{where }{{\mathpzc{Q}}}({{z}})&{:=}&-{{z}}-{{z}}^{-1}-\beta^{-1}{\mathcal{E}}_{{\upkappa}}, \label{q2_sq}\\
{{\lambda}}&{:=}&\frac{{{\mathpzc{r}}}-{{\mathpzc{h}}}}{{{\mathpzc{r}}}+{{\mathpzc{h}}}}, {{\mathpzc{h}}}{:=}\sqrt{{\mathpzc{H}}}, {{\mathpzc{r}}}{:=}\sqrt{{\mathpzc{R}}}, \label{lambdadef_sq}\\
{{\mathpzc{H}}}&{:=}&{\mathpzc{Q}}-2, {{\mathpzc{R}}}{:=} {\mathpzc{Q}}+2.
\label{HR_sq}
\end{eqnarray}\label{dHelmholtzF_sq_full}\end{subequations}
The complex functions ${\mathpzc{h}}, {\mathpzc{r}}, {{\mathpzc{H}}}$, ${{\mathpzc{R}}}$, and ${\lambda}$ are defined on ${\mathbb{C}}\setminus{\mathscr{B}}$ where ${\mathscr{B}}$ denotes the union of branch cuts for ${{\lambda}}$, borne out of the chosen branch
$-\pi<\arg {{\mathpzc{H}}}({{z}}) <\pi, {\mathbb{R}}e {{\mathpzc{h}}}({{z}})>0, {\mathbb{R}}e {{\mathpzc{r}}}({{z}})>0, {\text{\rm sgn}} \Im {{\mathpzc{h}}}({{z}})={\text{\rm sgn}} \Im {{\mathpzc{r}}}({{z}}), $
for ${{\mathpzc{h}}}$ and ${{\mathpzc{r}}}$ such that 
$|{{\lambda}}({{z}})|\le1, {{z}}\in{\mathbb{C}}\setminus{\mathscr{B}},$ as $\Im{\mathcal{E}}_{{\upkappa}}$ in \eqref{lambdadef_sq} is positive. 
The general solution of \eqref{dHelmholtzF_sq} is given by the expression
\begin{equation}\begin{split}
{\su}^{{\mathit{s}}}_{{\mathtt{y}}}{}^F={\mathit{c}}_1 {\lambda}^{{\mathtt{y}}}+{\mathit{c}}_2 {\lambda}^{-{\mathtt{y}}}, 
\label{gensol_sq}
\end{split}\end{equation}
where ${\mathit{c}}_{1, 2}$ are arbitrary analytic functions of ${z}$ in ${{\mathscr{A}}}$ (to be specified later).

\subsection{}
\label{appRecall_tg}
Following \cite{Bls4}, using the discrete Fourier transform defined by \eqref{unpm}, the discrete Helmholtz equation \eqref{dHelmholtz_tg} can be expressed as \eqref{dHelmholtzF_sq} for all ${\mathtt{y}}\in{\mathbb{Z}}$ inside the lattice but away from the boundary,
where
\begin{equation}\begin{split}
{{\mathpzc{Q}}}({{z}}){:=}\frac{\frac{3}{2}-{{z}}^2-{{z}}^{-2}-\frac{3}{2}\beta^{-1}{\mathcal{E}}_{{\upkappa}}}{{{z}}+{{z}}^{-1}}, {{z}}\in{\mathbb{C}}.
\label{q2_tg}
\end{split}\end{equation}
The general solution 
is again given by the expression \eqref{gensol_sq} (using \eqref{lambdadef_sq})
but with ${\mathpzc{Q}}$ given by \eqref{q2_tg}.
The zeros of ${{\mathpzc{H}}}$ are ${{z}}_{{\mathpzc{h}}{}}, {{z}}_{{\mathpzc{h}}{a}}, 1/{{z}}_{{\mathpzc{h}}{}},$ and $1/{{z}}_{{\mathpzc{h}}{a}}$, where
\begin{subequations}\begin{eqnarray}
{{z}}_{{\mathpzc{h}}{}}=\frac{1}{4}(-2+\sqrt{6} \sqrt{3-\beta^{-1}{\mathcal{E}}_{{\upkappa}}}\notag\\
-\sqrt{6(1-\beta^{-1}{\mathcal{E}}_{{\upkappa}})-4\sqrt{6} \sqrt{3-\beta^{-1}{\mathcal{E}}_{{\upkappa}}}}),\label{h01}\\
\text{and }
{{z}}_{{\mathpzc{h}} a}=\frac{1}{4}(-2-\sqrt{6} \sqrt{3-\beta^{-1}{\mathcal{E}}_{{\upkappa}}}\notag\\
+\sqrt{6(1-\beta^{-1}{\mathcal{E}}_{{\upkappa}})+4\sqrt{6} \sqrt{3-\beta^{-1}{\mathcal{E}}_{{\upkappa}}}}),\label{h02}
\end{eqnarray}\label{h0}\end{subequations}
and these four points in ${\mathbb{C}}$ are also the branch points of ${{\mathpzc{h}}}$. Due to the property ${{\mathpzc{R}}}({{z}})=-{{\mathpzc{H}}}(-{{z}}),$ the zeros of ${{\mathpzc{R}}}$ are related to those of ${{\mathpzc{H}}}$ through multiplication by $-1$. The zeros of ${{\mathpzc{R}}}$ are ${{z}}_{{\mathpzc{r}}{}}(=-{{z}}_{{\mathpzc{h}}{}}), {{z}}_{{\mathpzc{r}}{a}}(=-{{z}}_{{\mathpzc{h}}{a}}), 1/{{z}}_{{\mathpzc{r}}{}}(=-1/{{z}}_{{\mathpzc{h}}{}}),$ and $1/{{z}}_{{\mathpzc{r}}{a}}(=-1/{{z}}_{{\mathpzc{h}}{a}})$ and these four points are also the branch points of ${{\mathpzc{r}}}$. 
Note that the square root in the expression for ${{z}}_{{\mathpzc{h}}{}}$ and ${{z}}_{{\mathpzc{h}}{a}}$ 
is chosen such that ${{z}}_{{\mathpzc{h}}{}}$ and ${{z}}_{{\mathpzc{h}}{a}}$ lie inside the unit circle ${\mathbb{T}}$ (since it is assumed that $\Im{\mathcal{E}}_{{\upkappa}}>0$).

\section{Multiplicative factorization of kernel}
\label{app_multfac}
The multiplicative factorization of ${{\mathpzc{L}}}$ is \cite{Noble} 
\begin{subequations}
\begin{equation}\begin{split}
{\mathpzc{L}}({{z}})={\mathpzc{L}}_{+}({{z}}){\mathpzc{L}}_{-}({{z}}), {{z}}\in{{\mathscr{A}}}_{L}, 
\label{Lpmexp}
\end{split}\end{equation}
where the factors ${\mathpzc{L}}_{\pm}$ are given by 
\begin{equation}\begin{split}
{\mathpzc{L}}_{\pm} ({{z}})=\exp(\pm\frac{1}{2\pi i} \oint_{{{\mathcal{C}}}}\frac{\log {\mathpzc{L}}({\zeta})}{{{z}}-{\zeta}}d{\zeta}), {{z}}\in{\mathbb{C}}\\
\text{ such that }|{{z}}|\gtrless {{\mathit{R}}}_{L}^{\pm1}.
\label{Lpm}
\end{split}\end{equation} 
\end{subequations}
In \eqref{Lpm}, ${{\mathcal{C}}}$ is any rectifiable, closed, counterclockwise contour lying in annulus of analyticity ${{\mathscr{A}}}$ for ${\mathpzc{L}}$. 
Also it is implicitly assumed that ${\mathpzc{L}}_{\pm}({{z}})={\mathpzc{L}}_{\mp}({{z}}^{-1})$, 
allowing the representation to be unique \cite{Noble}. 
Notice that the function ${{\mathpzc{L}}}_{+}$ (resp. ${{\mathpzc{L}}}_{-}$) is analytic, in fact it has neither poles nor zeros, in the exterior (resp. interior) of a disk centered at $0$ in ${\mathbb{C}}$ with radius ${{\mathit{R}}}_{L}$ (resp. ${{\mathit{R}}}_{L}^{-1}$). 
\section{Auxiliary details for the geometric part of the solution} 
\label{appAuxgeosol}
The geometric part of the solution modulo the incident wave \eqref{uinc_sq} can be constructed using the incident wave and reflected wave in the `bulk' lattice 
\begin{equation}\begin{split}
{\su}_{{\mathtt{x}}, {\mathtt{y}}}^{{\mathrm{inc}}{r}} ={\su}_{{\mathtt{x}}, {\mathtt{y}}}^{\mathrm{inc}}+c{\su}^{{r}}_{{\mathtt{x}}, {\mathtt{y}}}, {\mathtt{x}}\in{\mathbb{Z}}, {\mathtt{y}}\in{\mathbb{Z}^+},
\end{split}\end{equation}
where $c$ is needed to be determined. 
Since, it is required that ${\su}_{{\mathtt{x}}, {\mathtt{y}}}^{{\mathrm{inc}}{r}}=0$ at ${\mathtt{y}}=1, {\mathtt{x}}<0,$ or ${\mathtt{y}}=0, {\mathtt{x}}\ge0$,
${\su}^{{r}}_{\cdot, 0}=-{\su}_{\cdot, 0}^{\mathrm{inc}}$. 
Thus,
\begin{equation}\begin{split}{\su}_{{\mathtt{x}}, {\mathtt{y}}}^{{\mathrm{inc}}{r}}({s}) 
={{\mathrm{A}}}e^{-i{\upkappa}_x {\mathtt{x}}-i{\upkappa}_y {\mathtt{y}}}-{{\mathrm{A}}}
\begin{cases}
e^{-i{\upkappa}_x {\mathtt{x}}+i{\upkappa}_y {\mathtt{y}}}\text{ for }{s}={A},\\
e^{-i{\upkappa}_x {\mathtt{x}}+i{\upkappa}_y ({\mathtt{y}}-2)}\text{ for }{s}={B},\\
\end{cases}
\label{uincref_sq}
\end{split}\end{equation}
for all ${\mathtt{x}}\in{\mathbb{Z}}, {\mathtt{y}}\in{\mathbb{Z}^+},$ where ${s}={A}$ corresponds to the right side of step while ${s}={B}$ corresponds to the left side of step. Suppose ${\theta}={\theta}_{{r}}$ corresponds to the angle of ray emanating from the reflected waves from the right side of the step. Then, it is easy to see that
\begin{equation}\begin{split}
{\su}_{{\mathtt{x}}, {\mathtt{y}}}^{{g}}={\su}_{{\mathtt{x}}, {\mathtt{y}}}^{{\mathrm{inc}}{r}}({{A}}){{{\mathit{H}}}({\theta}_{{r}}-{\theta})}+{\su}_{{\mathtt{x}}, {\mathtt{y}}}^{{\mathrm{inc}}{r}}({{B}}){{{\mathit{H}}}({\theta}-{\theta}_{{r}})}.
\end{split}\end{equation}
Note that 
${\su}_{{\mathtt{x}}, {\mathtt{y}}}^{{g}}=\sum_{{s}={A},{B}}{\su}^{{\mathit{s}}}_{{\mathtt{x}}, {\mathtt{y}}}|_{{P}{s}}$
as the second term that appears in the far-field approximation \eqref{uapprox}.

\section{Auxiliary details for the exact solution} 
\label{appAuxexactsol}
\subsection{}
\label{appAuxexactsol_sq}

It follows from \eqref{uinc_sq}, \eqref{WHCeq_sq_X_bulkinc}, after applying the multiplicative factorization ${{{\mathpzc{L}}}}={{{\mathpzc{L}}}}_{+}{{{\mathpzc{L}}}}_{-}$, 
that ${{{\mathpzc{L}}}}_{+}{\su}^{{\mathit{s}}}_{2;+}+{{{\mathpzc{L}}}}_{-}^{-1}{\su}^{{\mathit{s}}}_{2;-}={{\mathpzc{C}}}({{z}}), 
{{z}}\in{{\mathscr{A}}}$ where
\begin{subequations}
\begin{equation}\begin{split}
{{\mathpzc{C}}}({{z}})=({{{\mathpzc{L}}}}_{-}^{-1}-{{{\mathpzc{L}}}}_{+})({{\mathpzc{w}}}-{{\mathpzc{Q}}}{{\mathrm{A}}}e^{-i{\upkappa}_y}\delta_{D-}({{z}}{{z}}_{{P}}^{-1}e^{{-}{\mathfrak{e}}})\\
-{{\mathrm{A}}}\delta_{D+}({{z}}{{z}}_{{P}}^{-1}e^{{+}{\mathfrak{e}}}), \quad {{z}}\in{{\mathscr{A}}}. \label{Cz_gh_sq}
\end{split}\end{equation}
Hence, 
\begin{equation}\begin{split}
{{\mathpzc{C}}}_\pm({{z}})=\mp{\su}^{{\mathit{s}}}_{-1, 1}{{{{\mathpzc{L}}}}}_{\pm}^{\pm1}({{z}})\pm{{z}} {\su}^{{\mathit{s}}}_{0, 1}({{{{\mathpzc{L}}}}}_{\pm}^{\pm1}({{z}})-{l}_{+0})\\
\pm{{\mathrm{A}}}e^{-i{\upkappa}_y}\delta_{D-}({{z}} {{z}}_{{P}}^{-1}e^{{-}{\mathfrak{e}}})\\
\big({{\mathpzc{Q}}}({{z}}){{{{\mathpzc{L}}}}}_{\pm}^{\pm1}({{z}})-{{{{\mathpzc{L}}}}}_+({{z}}_{{P}}){{\mathpzc{Q}}}({{z}}_{{P}})+{l}_{+0}({{z}}-{{z}}_{{P}})\big)\\
\mp({{{\mathpzc{L}}}}_{-}^{-1}({{z}}_{{P}})-{{{\mathpzc{L}}}}_{\pm}^{\pm1}({{z}})){{\mathrm{A}}}\delta_{D+}({{z}}{{z}}_{{P}}^{-1}e^{{+}{\mathfrak{e}}}), 
\label{Cpm_gh_sq}
\end{split}\end{equation}
\end{subequations}
where ${l}_{+0}=\lim_{{z}\to\infty}{{{{\mathpzc{L}}}}}_+({{z}}).$ 
In \eqref{Cpm_gh_sq}, $\pm$ signs concur. 
After an application of the Liouville theorem \cite{Noble,Bls0}, in terms of the one-sided discrete Fourier transform the complex function ${\su}^{{\mathit{s}}}_{2}{}^F$ is given by 
\begin{equation}\begin{split}{\su}^{{\mathit{s}}}_{2;\pm}({{z}})={{\mathpzc{C}}}_{\pm}({{z}}){{{\mathpzc{L}}}}_{\pm}^{\mp1}({{z}}), {{z}}\in{\mathbb{C}}, |{{z}}|\gtrless\bfrac{\max}{\min}\{{{\mathit{R}}}_{\pm}, {{\mathit{R}}}_{L_{{\mathscr{S}}}}^{\pm1}\}.\label{u2pmsol_gh_sq}\end{split}\end{equation}
Note that ${{z}}_{{\mathpzc{q}}}$ and ${{z}}_{{\mathpzc{q}}}^{-1}$ are the two zeros of ${\mathpzc{Q}}$ (with $|{{z}}_{{\mathpzc{q}}}|<1$), i.e., ${\mathpzc{Q}}({z})={z}_{{\mathpzc{q}}}^{-1}(1-{z}_{{\mathpzc{q}}}{z})(1-{z}_{{\mathpzc{q}}}{z}^{-1})=-{z}^{-1}({z}-{z}_{{\mathpzc{q}}})({z}-{z}_{{\mathpzc{q}}}^{-1})$; ${\mathpzc{Q}}_\pm({z})={z}_{{\mathpzc{q}}}^{-1/2}(1-{z}_{{\mathpzc{q}}}{z}^{\mp1})$.

\subsection{}
\label{appAuxexactsol_tg}
It follows from \eqref{uinc_sq}, \eqref{WHCeq_tg_X}, after applying the multiplicative factorization ${\mathpzc{L}}Ntg={\mathpzc{L}}Ntg_{+}{\mathpzc{L}}Ntg_{-}$, that $({{z}}+{{z}}^{-1}){\mathpzc{L}}Ntg_{+}{\su}^{{\mathit{s}}}_{2;+}+({{z}}+{{z}}^{-1}){\su}^{{\mathit{s}}}_{2;-}{\mathpzc{L}}Ntg_{-}^{-1}={{\mathpzc{C}}}({{z}}), 
{{z}}\in{{\mathscr{A}}}$ where
\begin{subequations}
\begin{equation}\begin{split}
{{\mathpzc{C}}}({{z}})&=({\mathpzc{L}}Ntg_{-}^{-1}-{{{\mathpzc{L}}}}_{+})({\mathpzc{w}}-(-{{z}} u^{\mathrm{inc}}_{0, 0}+u^{\mathrm{inc}}_{-1, 0})\\
&-({{z}}+{{z}}^{-1}){{\mathpzc{Q}}}({{z}}){{\mathrm{A}}}e^{-i{\upkappa}_y}\delta_{D-}({{z}}{{z}}_{{P}}^{-1}e^{{-}{\mathfrak{e}}})\\
-({{z}}+{{z}}^{-1}){{\mathrm{A}}}\delta_{D+}({{z}}{{z}}_{{P}}^{-1}e^{{+}{\mathfrak{e}}})), \quad {{z}}\in{{\mathscr{A}}}. \label{Cz_gh_tg}
\end{split}\end{equation}
Using the detailed expressions provided by \cite{Bls4} for the infinite lattice (the expression for ${\mathpzc{C}}_{\pm}$ can be also identified with (B.4) of \cite{Bls4}), as well as by \cite{Bls4} for the rigid constraint induced bifurcated waveguides' problem, 
\begin{equation}\begin{split}
{{\mathpzc{C}}}_\pm({{z}})=\mp{{z}}^2(-{\su}^{{\mathit{s}}}_{0, 1})({{{\mathpzc{L}}}}_\pm^{\pm1}({{z}})-{l}_{+0}-{l}_{+1}{{z}}^{-1})\\
\mp{{z}}(-{\su}^{{\mathit{s}}}_{1, 1}-{\su}^{{\mathit{s}}}_{0, 2})({{{\mathpzc{L}}}}_\pm^{\pm1}({{z}})-{l}_{+0})\\\mp(-{{\mathrm{A}}}e^{-i{\upkappa}_y}{{z}}_{{P}}^{-2}+{\su}^{{\mathit{s}}}_{-1, 2}){({{{\mathpzc{L}}}}_\pm^{\pm1}({{z}})-{\overline{l}}_{-0})}\\\mp{{z}}^{-1}(-{{\mathrm{A}}}e^{-i{\upkappa}_y}{{z}}_{{P}}^{-1})({{{\mathpzc{L}}}}_\pm^{\pm1}({{z}})-{\overline{l}}_{-0}{-{\overline{l}}_{-1}{z}})\\\pm{{{\mathrm{A}}}e^{-i{\upkappa}_y}}(\overline{l}_{-0}{({{z}}^{-1}+{{z}}_{{P}}^{-1})+\overline{l}_{-1}){{z}}_{{P}}^{-1}}\\\pm{{{\mathrm{A}}}e^{-i{\upkappa}_y}}\delta_{D-}({{z}} {{z}}_{{P}}^{-1}e^{{-}{\mathfrak{e}}})\big({{{\mathpzc{L}}}}_\pm^{\pm1}({{z}})({{z}}+{{z}}^{-1}){{\mathpzc{Q}}}({{z}})\\-{{{\mathpzc{L}}}}_+({{z}}_{{P}})({{z}}_{{P}}+{{z}}^{-1}_{{P}}){{\mathpzc{Q}}}({{z}}_{{P}})+{l}_{+0}({{z}}^2-{{z}}_{{P}}^2)
+{l}_{+1}({{z}}-{{z}}_{{P}})\big)\\
\mp({{z}}(-{\su}^{\mathrm{inc}}_{0, 2})({{{\mathpzc{L}}}}_\pm^{\pm1}({{z}})-{l}_{+0})+{\su}^{\mathrm{inc}}_{-1, 2}({{{\mathpzc{L}}}}_\pm^{\pm1}({{z}})-{\overline{l}}_{-0}))\\
\pm{{{\mathrm{A}}}}\delta_{D+}({{z}} {{z}}_{{P}}^{-1}e^{{+}{\mathfrak{e}}})\big({{{\mathpzc{L}}}}_\pm^{\pm1}({{z}})({{z}}+{{z}}^{-1})\\-{{{\mathpzc{L}}}}_-^{-1}({{z}}_{{P}})({{z}}_{{P}}+{{z}}^{-1}_{{P}})-{l}_{+0}({{z}}-{{z}}_{{P}})
-{\overline{l}}_{-0}({{z}}^{-1}-{{z}}_{{P}}^{-1})\big), 
\label{Cpm_gh_tg}
\end{split}\end{equation}
\end{subequations}
where ${l}_{+0}=\lim_{{{z}}\to\infty}{{{{\mathpzc{L}}}}}_+({{z}})$, ${l}_{+1}=\lim_{{{z}}\to\infty}{{z}}({{{{\mathpzc{L}}}}}_+({{z}})-{l}_{+0}), {\overline{l}}_{-0}=\lim_{{{z}}\to0}{{{{\mathpzc{L}}}}}^{-1}_-({{z}})$, etc. Then
\begin{equation}\begin{split}J({z})=&(({{z}}+{{z}}^{-1}){{{\mathpzc{L}}}}_+({{z}}){\su}^{{\mathit{s}}}_{2;+}({{z}})+{\overline{l}}_{-0}{\su}^{{\mathit{s}}}_{-1,2}\\&-{{z}} {l}_{+0}{\su}^{{\mathit{s}}}_{0,2})-{{\mathpzc{C}}}_+({{z}})\\=&-\frac{{{z}}+{{z}}^{-1}}{{{{\mathpzc{L}}}}_-({{z}})}{\su}^{{\mathit{s}}}_{2;-}({{z}})+{{\mathpzc{C}}}_-({{z}})+{\overline{l}}_{-0}{\su}^{{\mathit{s}}}_{-1,2}\\&-{{z}} {l}_{+0}{\su}^{{\mathit{s}}}_{0,2} \text{ on }{{\mathscr{A}}}, 
\label{Jc_cd}
\end{split}\end{equation}
holds. Notice that as ${z}\to\infty,$ $J({z})\sim$ constant, on the other hand, as ${z}\to0,$ $J({z})\sim{\mathpzc{C}}_-(0)$, where ${\mathpzc{C}}_-(0)=0$.
The function ${{\mathpzc{C}}}_+({{z}})$ (resp. ${{\mathpzc{C}}}_-({{z}})$) is analytic at ${{z}}\in{\mathbb{C}}$ such that $|{{z}}|>\max\{{{\mathit{R}}}_+, {{\mathit{R}}}_{{{{\mathpzc{L}}}}}\}$ (resp. $|{{z}}|<\min\{{{\mathit{R}}}_-, {{\mathit{R}}}_{{{{\mathpzc{L}}}}}^{-1}\}$). 
In \eqref{Cpm_gh_tg}, $\pm$ signs concur. 
By an application of
the Liouville's theorem, the solution of the discrete Wiener--Hopf equation \eqref{WHCeq_tg_X} can be written in terms of one-sided discrete Fourier transforms as
\begin{equation}\begin{split}
({{z}}+{{z}}^{-1}){\su}^{{\mathit{s}}}_{2;\pm}({{z}})&={{{\mathpzc{L}}}}_\pm^{\mp1}({{z}})(\pm{{z}}{l}_{+0}{\su}^{{\mathit{s}}}_{0, 2}{\mp{\overline{l}}_{-0}{\su}^{{\mathit{s}}}_{-1,2}}\\
&+ {{\mathpzc{C}}}_\pm({{z}})), {{z}}\in{\mathbb{C}}, |{{z}}|\gtrless\bfrac{\max}{\min}\{{{\mathit{R}}}_{\pm}, {{\mathit{R}}}_{L}^{\pm1}\}. 
\label{u2pmsol_gh_tg}
\end{split}\end{equation}
Substituting \eqref{u2pmsol_gh_tg} in \eqref{sliteq_tg_X}, the expression for ${\su}^{{\mathit{s}}}_{1;+}$, in terms of ${\su}^{{\mathit{s}}}_{0, 1}, {\su}^{{\mathit{s}}}_{1, 1}, {\su}^{{\mathit{s}}}_{0, 2}$ and ${\su}^{{\mathit{s}}}_{-1, 2}$, is found, 
it is further simplified to obtain ${\su}^{{}}_{0, 1}$ which is similar to the expression found by \cite{Bls4}. In particular, using the residue calculus, 
${\su}^{{}}_{0, 1}$ is obtained as 
\begin{subequations}
\begin{equation}\begin{split}
{\su}^{{}}_{0, 1}=&-{{\mathrm{A}}}e^{-i{\upkappa}_y}\frac{{{{{\mathpzc{L}}}}}_+({{z}}_{{P}})({{z}}_{{P}}+{{z}}^{-1}_{{P}}){{\mathpzc{Q}}}({{z}}_{{P}})}{(({{z}}_{{P}}+{{z}}_{{\mathpzc{q}}}^{-1}){l}_{+0}+{l}_{+1})({{z}}_{{P}}-{{z}}_{{\mathpzc{q}}}^{-1})}\\&
-{{{\mathrm{A}}}}\frac{1-{{z}}_{{\mathpzc{q}}}^{2}{{z}}_{{P}}^{-2}}{{\mathpzc{Q}}({{z}}_{{P}})}\frac{{{z}}_{{P}}+{{z}}_{{\mathpzc{q}}}^{-1}}{{{{\mathpzc{L}}}}_-({{z}}_{{P}})}\frac{1}{({{z}}_{{P}}+{{z}}_{{\mathpzc{q}}}^{-1}){l}_{+0}+{l}_{+1}}.
\end{split}\end{equation} 
Further, ${l}_{+1}=0$ due to the presence of even powers of ${{z}}$ in the kernel (an indirect consequence of the double degeneracy of energy band relation of ${\mathfrak{R}}$ since it is a union of two decoupled lattices ${\mathfrak{T}}$ and the replicated ${\mathfrak{T}}^{{^\mathrm{R}}}$). 
Hence,
\begin{equation}\begin{split}
{\su}^{{}}_{0, 1}
={{\mathrm{A}}}\frac{(1-{{z}}_{{\mathpzc{q}}}^{2}{{z}}_{{P}}^{-2})}{{l}_{+0}}\big(e^{-i{\upkappa}_y}{{{{\mathpzc{L}}}}}_+({{z}}_{{P}})\\
-\frac{1}{{\mathpzc{Q}}({{z}}_{{P}})}{{{\mathpzc{L}}}}_-^{-1}({{z}}_{{P}})\big).
\label{u01tot}
\end{split}\end{equation} 
\end{subequations}
Above expression can also compared with mathematically analogous problems studied by \cite{Bls10mixed} for the square lattice half-planes with a different set of boundary conditions. 
By translational symmetry ${\su}^{{}}_{1, 1}=e^{-i{\upkappa}_x}{\su}^{{}}_{0, 1}$, which is plotted in Fig. \ref{Re_Im_Abs_u11_tot_tg}.
Note that
$({{z}}+{{z}}^{-1}){{\mathpzc{Q}}}({{z}})=-{{z}}^{-2}({{z}}^2-{{z}}^2_{{\mathpzc{q}}})({{z}}^2-{{z}}_{{\mathpzc{q}}}^{-2}),$ so that $(({{z}}+{{z}}^{-1}){\mathpzc{Q}})_\pm({z})=(-{z}_{{\mathpzc{q}}})^{-1/2}{z}_{{\mathpzc{q}}}^{-1/2}(1-{z}_{{\mathpzc{q}}}^2{z}^{\mp2})$.

\section{Auxiliary details for far-field approximation} 
\label{appAuxdetail}
\subsection{}
\label{appAuxdetail_sq}
Following \cite{Bls0},
let ${\upxi}_{\bfrac{i}{f}}=\mp\pi+\pi {\mathit{H}}(\beta^{-1}{\mathcal{E}}_{{\upkappa}}){\mathit{H}}(4-\beta^{-1}{\mathcal{E}}_{{\upkappa}}).$ Let the ``polar'' coordinates $({{}R}, {\theta})$ for $({\mathtt{x}}, {\mathtt{y}})\in{{\mathbb{Z}^2}}$, be specified by the relations
\begin{equation}\begin{split}
{\mathtt{x}}={{}R}\cos{\theta}, {\mathtt{y}}={{}R}\sin{\theta},\\
{{}R}=\sqrt{{\mathtt{x}}^2+{\mathtt{y}}^2}>0, {\theta}\in[0, \pi).
\label{polar_sq}
\end{split}\end{equation}
The mapping ${{z}}=e^{-i{\upxi}}$ and polar coordinates $({{}R}, {\theta})$ are used to rewrite the solution \eqref{umnsol} as
\begin{equation}\begin{split}
{\su}^{{\mathit{s}}}_{{{\mathtt{x}}}, {{\mathtt{y}}}}=-\frac{1}{2\pi}{{\mathrm{A}}}{\mathtt{C}}_0\int_{{{\mathcal{C}}}_{{{\upxi}}}}\frac{{{\mathpzc{K}}}(e^{-i{{\upxi}}})e^{i{{}R}\upphi({{\upxi}})}}{e^{i({\upxi}-
{\upkappa}_x)}-1}e^{-i{\upeta}({\upxi})}d{{\upxi}},
\label{umnsolrt_sq}
\end{split}\end{equation}
where ${{\mathcal{C}}}_{{{\upxi}}}$ is a contour, traversed from ${\upxi}_i$ to ${\upxi}_f$, in the strip 
${{\mathscr{S}}}=\{{\upxi}\in{\mathbb{C}}: {\upxi}_1\in[{\upxi}_i, {\upxi}_f], -{\upkappa}_2<{\upxi}_2<{\upkappa}_2\cos{\Theta}\},$ and
\begin{equation}\begin{split}
\upphi({\upxi})={\upeta}({\upxi})\sin{\theta}-{\upxi}\cos{\theta}, \\
{\upeta}({\upxi})=-i\log{{\lambda}}(e^{-i{\upxi}}), {\upxi}\in {{\mathscr{S}}}. \label{phi_sq} 
\end{split}\end{equation}
The pole for the diffraction integral \eqref{umnsolrt_sq} (see \cite{Bls0} for details) 
is located on the contour of integration, ${{\mathcal{C}}}_{{\upxi}}$, at ${\upxi}=
{\upkappa}_x$ for all $\beta^{-1}{\mathcal{E}}_{{\upkappa}}\in(-4, 0)$, or for all $\beta^{-1}{\mathcal{E}}_{{\upkappa}}\in(0, 4)$ and admissible ${\Theta}\in[0, \pi/2]$, and at $
{\upkappa}_x+2\pi$ for all $\beta^{-1}{\mathcal{E}}_{{\upkappa}}\in(0, 4)$ and admissible ${\Theta}\in(\pi/2, \pi]$. {The function $\upphi$ \eqref{phi_sq} possesses a saddle point at ${\upxi}={\upxi}_{{S}}$ on ${\mathcal{C}}_{{\upxi}}$ if \cite{Erdelyi1,Felsen,Born} $\upphi'({\upxi}_{{S}})={\upeta}'({\upxi}_{{S}})\sin{\theta}-\cos{\theta}=0, \upphi''({\upxi}_{{S}})={\upeta}''({\upxi}_{{S}})\sin{\theta}\ne 0.$ The saddle point ${\upxi}_{{S}}$ for the diffraction integral \eqref{umnsolrt_sq} is same as that discussed in detail by \cite{Bls0}. 
} 
The criterion for non-zero contribution of pole of diffraction integral \eqref{umnsolrt_sq} can be translated into a criterion that involves ${\theta}$ and ${\Theta}$ using the fact that ${\upxi}_{{S}}$ varies monotonically with ${\theta}$ (Theorem 3.4, \cite{Bls0}). {See Fig. 5.1 in \cite{Bls0}, where an illustration is provided for some values of $\beta^{-1}{\mathcal{E}}_{{\upkappa}}$ of the dependence of ${\theta}_{{r}}$ on ${\Theta}$, i.e., ${\theta}$-${\Theta}$ relation corresponding to the coalescence of pole and saddle point of diffraction integral $
{\upkappa}_x={\upxi}_{{S}}$ for $\beta^{-1}{\mathcal{E}}_{{\upkappa}}\in(-4, 0)$ and $\beta^{-1}{\mathcal{E}}_{{\upkappa}}\in(0, 4)$. 

\subsection{}
\label{appAuxdetail_tg}
Suppose the polar coordinates $({{}R}, {\theta})$ 
are specified by the relations
\begin{equation}\begin{split}
{\mathtt{x}}=2{{}R}\cos{\theta}, {\mathtt{y}}
=\frac{2}{\sqrt{3}}{{}R}\sin{\theta}, \\
\text{ and }{{}R}=\sqrt{\frac{1}{4}{\mathtt{x}}^2+\frac{3}{4}{\mathtt{y}}^2}>0, {\theta}\in[0, \pi).
\label{polar_tg}
\end{split}\end{equation}

For $\beta^{-1}{\mathcal{E}}_{{\upkappa}}\in(-3, {7}/{{3}})
\cup({16}/{{3}}, {6})$, 
the analysis of asymptotic approximation \cite{Born} of the scattered wavefunction in far field 
follows \cite{Bls4}.
Let ${\upxi}_{\bfrac{i}{f}}=\mp\pi.$ 
Using ${{z}}=e^{-i{\upxi}}$ and the polar coordinates \eqref{polar_tg}, the solution 
\eqref{umnsol} is rewritten as \eqref{umnsolrt_sq}
where
\begin{equation}\begin{split}
\upphi({\upxi})=2(\frac{1}{\sqrt{3}}{\upeta}({\upxi})\sin{\theta}-{\upxi}\cos{\theta}), 
\label{phi_tg}
\end{split}\end{equation}
${\upeta}$ is given by \eqref{phi_sq}${}_2$ and ${{\mathcal{C}}}_{{{\upxi}}}$ is a contour, traversed from ${\upxi}_i$ to ${\upxi}_f$, in the strip 
${{\mathscr{S}}}=\{{\upxi}\in{\mathbb{C}}: {\upxi}_1\in[{\upxi}_i, {\upxi}_f], -{\frac{1}{2}}{\upkappa}_2<{\upxi}_2<{\frac{1}{2}}{\upkappa}_2\cos{\Theta}\}.$
The pole for the diffraction integral \eqref{umnsolrt_sq} 
is located close to the contour of integration, ${{\mathcal{C}}}_{{\upxi}}$, at ${\upxi}=
{\upkappa}_x$ for all $\beta^{-1}{\mathcal{E}}_{{\upkappa}}\in(-3, {7}/{{3}})$, or for all $\beta^{-1}{\mathcal{E}}_{{\upkappa}}\in({7}/{{3}}, {3})$ and admissible ${\Theta}\in[0, \pi]$. For the crack problem, there are two poles located at ${\upxi}=
{\upkappa}_x$ and ${\upxi}
=i\log (-e^{-i{\upkappa}_x})$. 

\end{document}